\documentclass[twocolumn]{aastex701}
\usepackage{xspace}
\usepackage{amsmath}
\usepackage{mhchem}
\usepackage[export]{adjustbox}

\newcommand{\Kepler}{Kepler\xspace}
\newcommand{\Ktwo}{K2\xspace}
\newcommand{\TESS}{TESS\xspace}
\newcommand{\Gaia}{Gaia\xspace}

\newcommand{\Exofast}{\texttt{EXOFASTv2}\xspace}
\newcommand{\exoplanet}{\texttt{exoplanet}\xspace}

\newcommand{\bjdtdb}{\ensuremath{\mathrm{BJD}_\mathrm{TDB}}\xspace}
\newcommand{\Rstar}{\ensuremath{R_{\star}}\xspace}

\newcommand{\rhostarc}{\ensuremath{\Tilde{\rho}_\star}\xspace}
\newcommand{\Rjup}{\ensuremath{R_\mathrm{J}}\xspace} 
\newcommand{\Mjup}{\ensuremath{M_\mathrm{J}}\xspace}
\newcommand{\Rearth}{\ensuremath{R_\oplus}\xspace} 
\newcommand{\Mearth}{\ensuremath{M_\oplus}\xspace}
\newcommand{\Rsun}{\ensuremath{R_\odot}\xspace} 
\newcommand{\Msun}{\ensuremath{M_\odot}\xspace}
\newcommand{\Rp}{\ensuremath{R_\mathrm{p}}\xspace}
\newcommand{\Mp}{\ensuremath{M_\mathrm{p}}\xspace}
\newcommand{\Mpsini}{\ensuremath{M_\mathrm{p}\sin{i}}\xspace}

\newcommand{\Teff}{\ensuremath{T_\mathrm{eff}}\xspace}
\newcommand{\logg}{\ensuremath{\log{g}}\xspace}
\newcommand{\feh}{\ensuremath{\mathrm{[Fe/H]}}\xspace}
\newcommand{\vsini}{\ensuremath{v\sin{i}}\xspace}

\newcommand{\Teq}{\ensuremath{T_\mathrm{eq}}\xspace}
\newcommand{\Prot}{\ensuremath{P_{\mathrm{rot}}}\xspace}
\newcommand{\Porb}{\ensuremath{P_{\mathrm{orb}}}\xspace}
\newcommand{\ror}{\ensuremath{\Rp/\Rstar}\xspace}

\newcommand{\ecosw}{\ensuremath{e\cos{\omega}}\xspace}
\newcommand{\esinw}{\ensuremath{e\sin{\omega}}\xspace}

\newcommand{\ms}{\ensuremath{\mathrm{m}\,\mathrm{s}^{-1}}\xspace}

\newcommand{\kms}{\ensuremath{\mathrm{km}\,\mathrm{s}^{-1}}\xspace}

\usepackage{afterpage}
\newenvironment{rotatepage}%
{\clearpage\pagebreak[4]\global\pdfpageattr\expandafter{\the\pdfpageattr/Rotate 90}}%
{\clearpage\pagebreak[4]\global\pdfpageattr\expandafter{\the\pdfpageattr/Rotate 0}}%

\graphicspath{{./}{figures/}}
\makeatletter
\def\input@path{{tables/}{figures/}}
\makeatother

\begin{document}

\title{TOI-3686: A Hot Jupiter and Sub-Neptune In A 2:1 Mean-Motion Resonance}

\author[0000-0001-7961-3907]{Samuel~W.~Yee}
\altaffiliation{51 Pegasi b Fellow}
\affiliation{Department of Physics \& Astronomy, University of California Los Angeles, Los Angeles, CA 90095, USA}
\affiliation{Center for Astrophysics \textbar Harvard \& Smithsonian, 60 Garden Street, Cambridge, MA 02138, USA}
\email[show]{syee@astro.ucla.edu}

\author{Tarisai~Dziire}
\affiliation{New York University Shanghai, 567 West Yangsi Rd, Pudong, Shanghai, 200124, China}
\affiliation{Center for Astrophysics \textbar Harvard \& Smithsonian, 60 Garden Street, Cambridge, MA 02138, USA}
\email{tid2005@nyu.edu}

\author[0000-0001-6588-9574]{Karen~A.~Collins}
\affiliation{Center for Astrophysics \textbar Harvard \& Smithsonian, 60 Garden Street, Cambridge, MA 02138, USA}
\email{karen.collins@cfa.harvard.edu}

\author[0000-0002-0514-5538]{Luke~G.~Bouma}
\affiliation{Caltech/IPAC, Pasadena, CA 91125, USA}
\email{lbouma@ipac.caltech.edu}

\author[0000-0002-9908-8705]{Daniel~Tamayo}
\affiliation{Department of Physics, Harvey Mudd College, 301 Platt Boulevard, Claremont, CA 91711, USA}
\email{dtamayo@g.hmc.edu}

\author[0000-0003-0967-2893]{Erik~A.~Petigura}
\email{petigura@astro.ucla.edu}
\affiliation{Department of Physics \& Astronomy, University of California Los Angeles, Los Angeles, CA 90095, USA}

\author[0000-0001-6637-5401]{Allyson~Bieryla}
\affiliation{Center for Astrophysics \textbar Harvard \& Smithsonian, 60 Garden Street, Cambridge, MA 02138, USA}
\email{abieryla@cfa.harvard.edu}

\author[0000-0001-9911-7388]{David W. Latham}
\affiliation{Center for Astrophysics \textbar Harvard \& Smithsonian, 60 Garden Street, Cambridge, MA 02138, USA}
\email{dlatham@cfa.harvard.edu}

\author[0000-0002-8964-8377]{Samuel~N.~Quinn}
\affiliation{Center for Astrophysics \textbar Harvard \& Smithsonian, 60 Garden Street, Cambridge, MA 02138, USA}
\email{squinn@cfa.harvard.edu}

\author[0000-0002-4265-047X]{Joshua N.\ Winn}
\affiliation{Department of Astrophysical Sciences, Princeton University, 4 Ivy Lane, Princeton, NJ 08544, USA}
\email{jnwinn@princeton.edu}

\author[0000-0001-8732-6166]{Joel D.\ Hartman}
\affiliation{Department of Astrophysical Sciences, Princeton University, 4 Ivy Lane, Princeton, NJ 08544, USA}
\email{jhartman@astro.princeton.edu}
\author[0000-0003-1713-3208]{Boris~S.~Safonov}			%
\affiliation{Sternberg Astronomical Institute, M.V. Lomonosov Moscow State University, 13, Universitetskij pr., 119234, Moscow, Russia}
\email{safonov10@gmail.com}
\author[0000-0002-6482-2180]{Raquel For\'{e}s-Toribio}
\affiliation{Department of Astronomy, The Ohio State University, 140 West 18th Avenue, Columbus, OH 43210, USA}
\affiliation{Center for Cosmology and Astroparticle Physics, The Ohio State University, 191 W. Woodruff Avenue, Columbus, OH 43210, USA}
\email{forestoribio.1@osu.edu}
\author[0000-0001-9833-2959]{Jose A. Mu\~noz}
\affiliation{Departamento de Astronom\'{\i}a y Astrof\'{\i}sica, Universidad de Valencia, E-46100 Burjassot, Valencia, Spain}
\affiliation{Observatorio Astron\'omico, Universidad de Valencia, E-46980 Paterna, Valencia, Spain}
\email{jmunoz@uv.es}
\author[0000-0003-1464-9276]{Khalid~Barkaoui}
\affiliation{Instituto de Astrof\'isica de Canarias (IAC), E-38205 La Laguna, Tenerife, Spain}
\affiliation{Astrobiology Research Unit, University of Li\`ege, All\'ee du 6 ao\^ut, 19, 4000 Li\`ege (Sart-Tilman), Belgium}
\affiliation{Department of Earth, Atmospheric and Planetary Sciences, Massachusetts Institute of Technology, Cambridge, MA 02139, USA}
\email{Khalid.Barkaoui@uliege.be}
\author[0000-0001-8227-1020]{Richard P.\ Schwarz}
\affiliation{Center for Astrophysics \textbar Harvard \& Smithsonian, 60 Garden Street, Cambridge, MA 02138, USA}
\email{rpschwarz@comcast.net}
\author[0000-0003-2127-8952]{Francis P. Wilkin} 
\affiliation{Department of Physics and Astronomy, Union College, 807 Union St., Schenectady, NY 12308, USA}
\email{wilkinf@union.edu}
\author[0000-0001-6981-8722]{Paul~Benni}
\affiliation{Acton Sky Portal (Private Observatory), Acton, MA, USA}
\email{pbenni@verizon.net}
\author[0000-0003-4647-7114]{Krzysztof Bernacki}
\affiliation{Silesian University of Technology, Department of Electronics, Electrical Engineering and Microelectronics, Akademicka 16, 44-100 GLiwice, Poland}
\email{Krzysztof.Bernacki@polsl.pl}
\author[0000-0002-4070-7831]{Nikita Chazov} 
\affiliation{Kourovka observatory, Ural Federal University, 19 Mira street, Yekaterinburg, Russia}
\email{nikita.googlov@gmail.com}
\author[0000-0002-9867-7938]{Eric G. Hintz}
\affiliation{Brigham Young University, N486 ESC, Provo, UT, 84602, USA}
\email{hintz@byu.edu}
\author[0009-0008-1272-9081]{Glauk Hizmo}
\affiliation{Department of Physics and Astronomy, Union College, 807 Union St., Schenectady, NY 12308, USA}
\email{glauk.hizmo@icloud.com}
\author[0009-0002-7986-9106]{George Iwaszek}
\affiliation{Nube Blanca Observatory, Berthoud, CO, USA}
\affiliation{American Association of Variable Star Observers, 49 Bay State Road, Cambridge, MA 02138, USA}
\email{giwaszek@gmail.com}
\author[0000-0003-3184-5228]{Adam Popowicz}
\affiliation{Silesian University of Technology, Department of Electronics, Electrical Engineering and Microelectronics, Akademicka 16, 44-100 GLiwice, Poland}
\email{apopowicz@polsl.pl}
\author[0000-0002-2190-3319]{Manfred Raetz} 
\affiliation{Private Observatory Herges-Hallenberg, Stiller Berg 6, 98587 Steinbach-Hallenberg, Germany}
\affiliation{Bundesdeutsche Arbeitsgemeinschaft f\"ur Ver\"anderliche Sterne e.V. (BAV), Munsterdamm 90, 12169 Berlin, Germany}
\email{mraetz.herges@stiller-berg.de}
\author{Miquel Serra-Ricart}
\affiliation{Instituto de Astrof\'isica de Canarias (IAC), E-38205 La Laguna, Tenerife, Spain}
\affiliation{Departamento de Astrof\'isica, Universidad de La Laguna (ULL), E-38206 La Laguna, Tenerife, Spain}
\affiliation{Light Bridges, SL. Observatorio Astron\'omico del Teide}
\email{miquel@lightbridges.es}
\author{Gregor Srdoc}
\affiliation{Kotizarovci Observatory, Sarsoni 90, 51216 Viskovo, Croatia}
\email{gregorsrdoc@gmail.com}
\author[0000-0003-4658-7567]{Denise C.\ Stephens} 
\affiliation{Brigham Young University, N486 ESC, Provo, UT, 84602, USA}
\email{denise_stephens@byu.edu}
\author{Joseph Williams}
\affiliation{Brigham Young University, N486 ESC, Provo, UT, 84602, USA}
\email{josephbwilliams2019@gmail.com}
\author[0000-0003-4590-0136]{Valerio Bozza}
\affiliation{Dipartimento di Fisica ``E.R. Caianiello'', Universit\`a di Salerno, Via Giovanni Paolo II 132, 84084 Fisciano, Italy}
\affiliation{Istituto Nazionale di Fisica Nucleare, Sezione di Napoli, Via Cintia, 80126 Napoli, Italy}
\email{valboz@sa.infn.it}
\author{Otabek Burkhonov}
\affiliation{Ulugh Beg Astronomical Institute, Uzbekistan Academy of Sciences, Tashkent, Uzbekistan}
\email{burkota75@gmail.com}
\author{Roger~Dymock}
\affiliation{British Astronomical Association, PO Box 702, Tonbridge, TN9 9TX, United Kingdom}
\email{	
roger.dymock@zohomail.eu}
\author{Salokhiddin Sh. Ergashev}
\affiliation{Samarkand State University named after Sh. Rashidov, University bulvar 15., Samarkand, Uzbekistan}
\email{ergashevsalohiddin111@gmail.com}
\author[0009-0009-8546-0703]{Karen Fischer}
\affiliation{Unistellar Citizen Astronomer, Placitas, NM}
\email{fischer7of9@hotmail.com}
\author[0000-0001-8621-6731]{Cristilyn N. Gardner-Watkins}
\affiliation{Center for Astrophysics \textbar Harvard \& Smithsonian, 60 Garden Street, Cambridge, MA 02138, USA}
\email{cristilynwatkins@gmail.com}
\author[0000-0001-9927-7269]{Ferran Grau Horta}
\affiliation{Observatori de Ca l'Ou, Carrer de dalt 18, Sant Martí Sesgueioles 08282, GEECAT, Barcelona, Spain}
\email{ferran.grau@gmail.com}
\author[0009-0009-2881-7112]{Adam Lark}
 \affiliation{Hamilton College, 198 College Hill Rd, Clinton, NY 13413, USA}
\email{alark@hamilton.edu}
\author[orcid=0009-0009-7566-8420,sname='Linnenkohl']{Katherine Linnenkohl}
\affiliation{Department of Physics \& Astronomy, University of Oregon, Eugene, OR 97403, USA}
\email{klinnenk@uoregon.edu>}
\author[0009-0007-4476-2831]{Ken Mason}
\affiliation{American Association of Variable Star Observers, 49 Bay State Road, Cambridge, MA 02138, USA}
\email{kenmason8317@gmail.com}
\author{Edward J. Michaels}
\affiliation{Waffelow Creek Observatory, 10780 FM 1878, Nacogdoches, TX 75961, USA}
\email{astroed@ejmj.net}
\author[orcid=0009-0005-5648-7107,sname='Mitchem']{Owen A.J. Mitchem}
\affiliation{Department of Physics \& Astronomy, University of Oregon, Eugene, OR 97403, USA}
\email{omitchem@uoregon.edu}
\author[0009-0002-5722-4024]{Seth M. Pritchard}
\affiliation{Citizen Scientist, Red Lion, PA}
\email{seth.p@live.com}
\author[0000-0001-6629-5399]{Lauren A.\ Sgro}
\affiliation{SETI Institute, Carl Sagan Center, 339 N. Bernardo Avenue, Suite 200, Mountain View, CA, USA}
\email{lsgro@seti.org}
\author{Neil B.\ Thomas}
\affiliation{Department of Astronautics, United States Air Force Academy, CO 80840, USA}
\email{nthomas819@aol.com}
\author{Siegfried Vanaverbeke} 
\affiliation{Vereniging Voor Sterrenkunde, Oude Bleken 12, 2400 Mol, Belgium}
\affiliation{Centre for Mathematical Plasma Astrophysics, Katholieke Universiteit Leuven,
Celestijnenlaan 200B, bus 2400, 3001 Leuven, Belgium}
\affiliation{Astrolab IRIS, Verbrandemolenstraat 5, Ypres, Belgium}
\email{siegfriedvanaverbeke@gmail.com}
\author[0000-0001-9592-5310]{Pieter Vuylsteke}
\affiliation{KU Leuven, Department of Physics and Astronomy, Institute of Astronomy, Celestijnenlaan 200d – box 2401, 3001 Leuven, Belgium}
\affiliation{Vereniging Voor Sterrenkunde, Oude Bleken 12, 2400 Mol, Belgium}
\email{pieter.vuylsteke@gmail.com}
\author[0000-0002-4371-3460]{Jeroen~Audenaert}
\affiliation{Department of Physics and Kavli Institute for Astrophysics and Space Research, Massachusetts Institute of Technology, Cambridge, MA 02139, USA}
\email{jeroena@mit.edu}
\author[0000-0003-1728-0304]{Keith Horne}
\affiliation{SUPA Physics and Astronomy, University of St. Andrews, Fife, KY16 9SS Scotland, UK}
\email{kdh1@st-andrews.ac.uk}
\author[0000-0001-9087-1245]{Felipe Murgas}
\affiliation{Instituto de Astrof\'isica de Canarias (IAC), E-38205 La Laguna, Tenerife, Spain}
\affiliation{Departamento de Astrof\'isica, Universidad de La Laguna (ULL), E-38206 La Laguna, Tenerife, Spain}
\email{fmurgas@iac.es}
\author[0000-0003-0571-2245]{Emma Nabbie}
\affiliation{Center for Astrophysics \textbar Harvard \& Smithsonian, 60 Garden Street, Cambridge, MA 02138, USA}
\email{emma.nabbie@cfa.harvard.edu}
\author[0000-0003-0987-1593]{Enric Palle}
\affiliation{Instituto de Astrof\'isica de Canarias (IAC), E-38205 La Laguna, Tenerife, Spain}
\affiliation{Departamento de Astrof\'isica, Universidad de La Laguna (ULL), E-38206 La Laguna, Tenerife, Spain}
\email{epalle@iac.es}
\author[0000-0003-3904-6754]{Ramotholo Sefako}  
\affiliation{South African Astronomical Observatory, P.O. Box 9, Observatory, Cape Town 7935, South Africa}
\email{rrs@saao.ac.za}
\author[0000-0002-1836-3120]{Avi Shporer}
\affiliation{Department of Physics and Kavli Institute for Astrophysics and Space Research, Massachusetts Institute of Technology, Cambridge, MA 02139, USA}
\email{shporer@mit.edu}
\begin{abstract}
Hot Jupiters are typically thought to be ``lonely'', but a small and growing subset of them have been found to have nearby planets in the same system. We present the discovery of a new member of this class of systems, TOI-3686. TOI-3686 c is a hot Jupiter with \Porb = 5.078 days detected by NASA's TESS mission that shows peak-to-peak transit timing variations (TTVs) of 40 minutes, amongst the largest ever observed for a hot Jupiter. A subsequent search of the TESS data revealed the transits of a second planet, TOI-3686 b, a sub-Neptune with \Porb = 2.54 days and peak-to-peak TTVs of 10 hours, $\sim 16\%$ of its orbital period.
We characterized the system through photodynamical modeling of photometry from TESS and ground-based facilities, as well as radial velocity measurements from the NEID and MAROON-X spectrographs.
We find that the two planets in the TOI-3686 system are deep in a 2:1 mean-motion resonance, making this the only hot Jupiter system currently known to be in a truly resonant configuration, which may be related to its $\lesssim1$~Gyr system age.
Systems like TOI-3686 likely formed through a dynamically quiescent pathway and were captured into resonance after migration through a disk, and may represent the high-mass end of processes that occur in compact systems of smaller planets.
\end{abstract}

\section{Introduction} 

Most hot Jupiters---giant planets with orbital periods of less than ten days---appear to be isolated, without nearby planetary companions \citep{Latham2011,Steffen2012,Huang2016,Sha2026}. 
This observation has been interpreted as evidence that hot Jupiters form farther out in the protoplanetary disk and arrive at their current close-in orbits via a dynamically disruptive, high-eccentricity migration pathway.
In this picture, a wide-orbiting giant planet is excited onto a highly eccentric orbit through mechanisms like planet-planet scattering \citep{Rasio1996,Weidenschilling1996} or secular interactions \citep{Wu2003,Fabrycky2007,Wu2011}, before undergoing orbit circularization due to tidal interactions with the host star \citep{Eggleton1998}.
Since any interior planets would likely be disrupted during the high-eccentricity phase \citep{Becker2026}, such hot Jupiters would end up isolated.

The existence of hot Jupiters with nearby planetary companions implies that not all of them form through high-eccentricity migration. Some may either form in-situ or migrate inward by dynamically quiescent means, such as disk-driven migration \citep[see, e.g., review by ][]{Dawson2018}.
The first example to be found was the WASP-47 system \citep{WASP47_Becker2015}, which contains a hot Jupiter flanked by two smaller planets that were detected by the \Ktwo mission. The prime \Kepler mission \citep{Kepler_Borucki2010} also discovered a hot Jupiter with an inner transiting companion, Kepler-730 \citep{Kepler730_Canas2019}.

However, hot Jupiters are intrinsically rare, occurring around $\sim 1\%$ of all Sun-like stars \citep{Wright2012}, so Kepler observed only a few dozen hot Jupiter systems.
NASA's ongoing \TESS mission \citep{TESS_Ricker15} is observing hundreds of hot Jupiters across the sky \citep{Yee2021b}, making it possible to find more transiting companions to hot Jupiters.
Indeed, \TESS has detected transiting planets both in previously known hot Jupiter systems (WASP-132, \citealt{WASP132_Hord2022}; WASP-84, \citealt{WASP84_Maciejewski2023}), as well as in newly discovered systems\citep[e.g.,][]{TOI1130_Huang2020b,TOI2000_Sha2023,TOI1408_Korth2024,TOI2494_TOI5143_Quinn2026,TOI1533_Mantovan2026}.

Apart from observing their transits directly, close-by companions to hot Jupiters may also be detected through their gravitational influence on the hot Jupiter, causing transit timing variations (TTVs).
Strong TTVs might especially be expected if hot Jupiters with companions arrive at their present locations through disk migration, which naturally captures neighboring planets into mean-motion resonances \citep{Nelson2002,Lee2002,Kley2004}, and resonant interactions lead to larger and more detectable TTVs \citep[e.g.,][]{Lithwick2012b,Nesvorny2016}.
Indeed, a few of the hot Jupiters with known transiting companions show TTVs, with semi-amplitudes ranging from $\sim 40$ seconds for WASP-47\,b \citep{WASP47_TTVs_Nascimbeni2023} to 8 minutes for TOI-1408\,b \citep{TOI1408_Korth2024} and 10 minutes for TOI-1130\,c \citep{TOI1130_Korth2023,TOI1130_Borsato2024}.

In this manuscript, we report the discovery and characterization of the TOI-3686 system, a new member of the class of hot Jupiters with nearby companions.
The system contains a hot Jupiter showing TTVs with a peak-to-peak amplitude of 40 minutes, some of the largest observed for a short-period giant planet.
We determined that these TTVs are due to a nearby sub-Neptune in a mean-motion resonance with the hot Jupiter.
We describe our observations in Section \ref{sec:observations}, stellar host characterization in Section \ref{sec:stellar_char}, and analysis of light curve data in Section \ref{sec:lc_analysis}.
We use $N$-body simulations to characterize the system in Section \ref{sec:photodynamical_analysis}, before discussing its dynamical state (Section \ref{sec:dynamics}) and its relation to the broader sample of hot Jupiters with companions (Section \ref{sec:context}).
We summarize in Section \ref{sec:conclusions}.

\section{Observations \label{sec:observations}}

\begin{figure*}
\includegraphics[width=\linewidth]{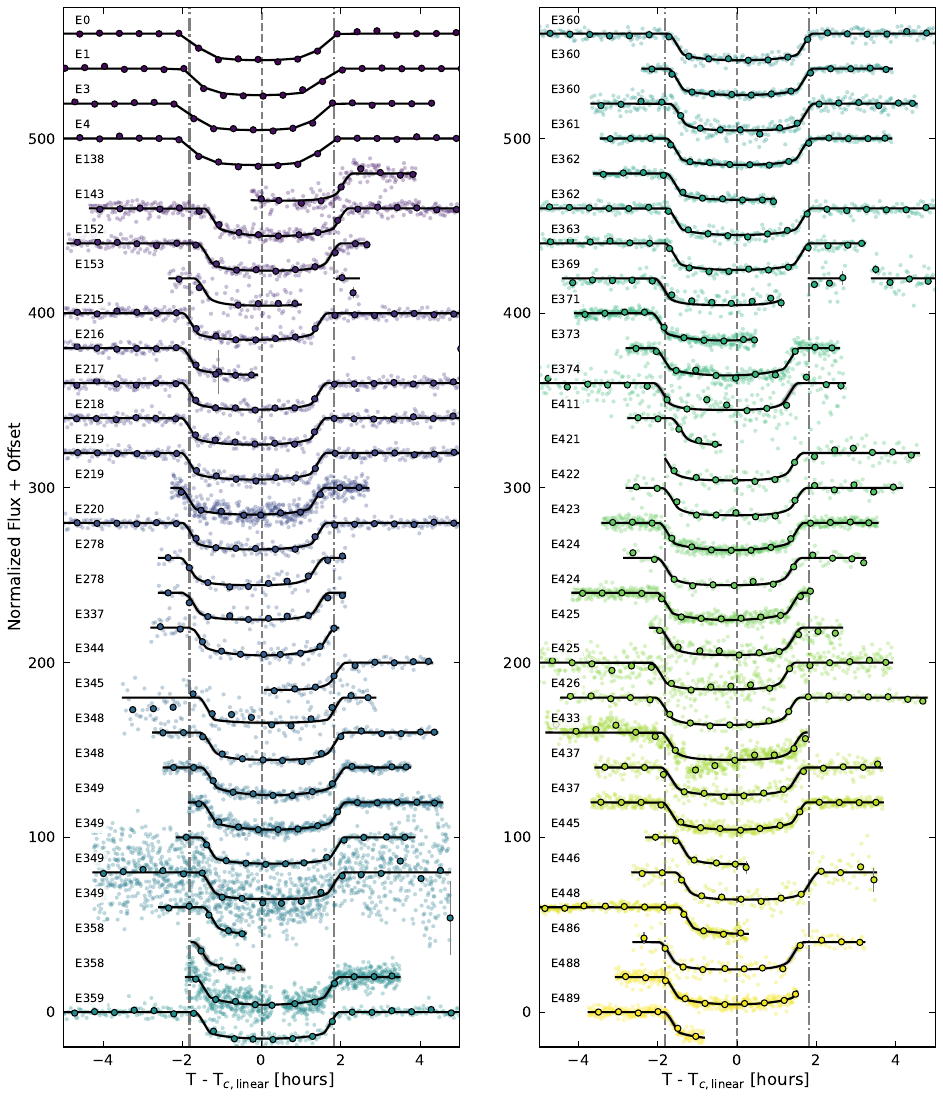}
\caption{TESS and ground-based transits of TOI-3686.01, phased to a linear orbital ephemeris. Small circles show the raw data, while large circles show data binned to 1800s cadence.
The data have been detrended by a model fit simultaneously with the transits.
Each transit is labeled with the epoch number (number of orbital periods since the first observed transit) and colored by time.
The black line shows the best-fit transit model.
Vertical dashed lines show the transit start, end, and mid-times expected for a constant period; the TTVs are visible as shifts of the observed ingress and egress relative to these. \label{fig:all_transits}}
\end{figure*}
\subsection{TESS Photometry}

TESS observed TOI-3686 in Sectors 18, 58, and 85, spanning a five year observation baseline from 2019 through 2024.
The TESS data show clear $1.5\%$ dips in the star's brightness occurring every 5.078 days (Figures \ref{fig:all_transits}, \ref{fig:tess_periodograms}).
The events in Sector 18 were alerted as a transiting planet candidate, TOI-3686.01, by the ``faint-star search'' \citep{TESS_QLP_Kunimoto2021} carried out for stars observed in the TESS Full Frame Images (FFIs), using light curves produced by the Quick Look Pipeline (QLP; \citealp{TESS_QLP_Huang2020a,TESS_QLP_Huang2020b}).
Following the identification of the transit events, the star was selected for higher cadence observations when it was again in TESS' field of view in Sectors 58 and 85.
We downloaded the TESS data for this target using the \texttt{lightkurve} package \citep{Lightkurve18}, selecting the light curves produced by the TESS-SPOC pipeline from the FFIs at 1800s cadence in Sector 18, and from the postage stamps at 120s cadence in Sectors 58 and 85 \citep{TESS_SPOC_Jenkins2016,TESS_SPOC_Caldwell2020}.

The TESS light curves of TOI-3686 also show longer term variability with a peak-to-peak amplitude of 1--2\% and which is quasi-periodic on a timescale of $\sim 5.55$~days (Figure \ref{fig:tess_periodograms}).
We attribute this photometric modulation to the rotation of starspots in and out of view, such that the periodicity is at the stellar rotation period, and discuss this signal further in Section \ref{ssec:youth}.

\subsection{Ground-Based Photometry}

We also obtained time-series transit photometry of TOI-3686.01 from telescopes on the ground.
These were made initially for false positive vetting, but once the presence of TTVs were noted, we began monitoring the transits more intensively to measure additional transit times.
Most of this follow-up was coordinated through the TESS Follow-up Observing Program Sub Group 1 (TFOP SG1; \citealt{TFOP_Collins2018}), which focuses on obtaining seeing-limited time-series photometry of TESS candidates. We used the {\tt TESS Transit Finder}, a customized version of the {\tt Tapir} software package \citep{TAPIR_Jensen2013}, to schedule the transit observations.
We obtained a total of 45 transit light curves from professional facilities, college observatories, as well as private observatories operated by citizen scientists across North America, Europe and Asia.
Table \ref{tab:transit_times} lists observation details and all facilities used.
Most of the ground-based lightcurves were extracted using {\tt AstroImageJ} \citep{AstroImageJ_Collins17}.

These observations resolved TOI-3686 from all nearby bright stars and confirmed that the transit events occur on-target with achromatic depths, favoring a planetary interpretation for the transits.
The geographic dispersion of the observers greatly extended the TTV baseline and improved phase coverage of the TTV curve compared with the TESS data alone.

\subsection{Speckle Imaging} \label{ssec:speckle_imaging}

To check for even closer stellar companions to TOI-3686, we also observed it with the Speckle Polarimeter \citep{SAI_Safonov2017,SAI_Strakhov2023} on the 2.5 m telescope at the Caucasian Mountain Observatory (CMO) of Sternberg Astronomical Institute (SAI) of Lomonosov Moscow State University. These observations achieved a limiting contrast ratio of $\Delta I = 5.2$~mag in $I$-band at a separation of $1^{\prime\prime}$, with no nearby sources detected down to instrumental limits.

We note that \textit{Gaia} DR3 \citep{GaiaDR3_Vallenari2022} lists a faint source (Gaia~DR3~465296935576885376, $G = 18.0$~mag) separated by $1\farcs66$ from TOI-3686. No astrometric data apart from position were available, so it is unclear whether this is a bound or background object. Given this source's faintness ($\Delta G = 6.24$~mag), it is impossible for it to be the source of either the transit events or long-term photometric modulation; thus, we do not consider this object further.

\subsection{Spectroscopic Observations}

We observed TOI-3686 with several high-resolution spectrographs, which we used to characterize the host star as well as measure precise radial velocities (RVs). All RV data are provided in a machine-readable table accompanying this paper.

\paragraph{TRES} We obtained 11 observations between Sep 2020 and Nov 2024  with the Tillinghast Reflector Echelle Spectrograph (TRES; \citealt{TRES_Furesz2008}) on the 1.5m telescope at the Fred Lawrence Whipple Observatory (FLWO). The spectra were reduced according to the pipeline described in \citet{TRES_Buchhave2010} and \citet{TRES_Quinn2012}. We derived relative RVs by first constructing a median-filtered template from all the observations and then cross-correlating this template against each individual observation. Long-term changes in the zero-point offset due to instrumental effects are corrected using standard star observations.

\paragraph{NEID} We obtained 41 spectra with the NEID spectrograph \citep{NEID_Schwab2016,NEID_Halverson2016} on the WIYN 3.5m telescope at Kitt Peak National Observatory (KPNO) between Oct 2021 and Mar 2025. We observed in High Resolution (HR) mode ($R \sim 110{,}000$) with exposure times of 300--900s. The data were reduced and RVs extracted using v1.4.2 of the NEID Data Reduction Pipeline.\footnote{\url{https://neid.ipac.caltech.edu/docs/NEID-DRP}}
The pipeline measures RVs by performing a cross-correlation between each observed spectrum and a weighted line mask derived for the star's spectral type \citep{Baranne1996,Pepe2002}.
These observations spanned several ``eras'' of NEID operations,\footnote{\url{https://neid.ipac.caltech.edu/docs/NEID-DRP/rveras.html}} between which offsets of a few \ms may occur, so we treat the different eras as having independent instrumental offsets.
In our subsequent analysis, we used only the 31 observations taken during the third RV era, as the earlier data were more sparsely sampled and had lower S/N.

\paragraph{MAROON-X} We also observed TOI-3686 with the MAROON-X spectrograph \citep{MAROONX_Seifahrt2016,MAROONX_Seifahrt2018,MAROONX_Seifahrt2020,MAROONX_Seifahrt2022} on the Gemini-North telescope (Prog. ID GN-2024B-FT-108; PI: Yee) on Maunakea, Hawaii, obtaining 23 epochs with 600s exposure times.
The data were reduced using the standard instrument pipeline.\footnote{\url{https://www.gemini.edu/instrumentation/maroon-x/data-reduction}}
Precise RVs were measured using a modified version of the \texttt{SERVAL} package \citep{SERVAL_Zechmeister2018}, which constructs an empirical co-added template from all observations, and then computes order-by-order shifts between each observation and the template through least squares fitting.
A combined RV for each of the blue and red channels of MAROON-X are derived from a weighted mean of the RVs across all orders in each arm.
To account for potential offsets between the two arms and between observing runs \citep{MAROONXOffsets_Basant2025}, we treated each combination of the above as separate instruments.

\subsubsection{Preliminary RV modeling \label{sssec:rv_analysis}}
We performed a preliminary fit of a single-planet Keplerian to the RV data, constrained to the best-fit linear ephemeris from the transit data.
This produced a reasonable fit to the data, with an RV semi-amplitude of  $K = 55.5^{+3.6}_{-3.3}\,\ms$, corresponding to a planetary mass of $\Mpsini = 0.48\pm0.04\,\Mjup$, although this measurement does not account for other sources of RV variation.
Together with the ground-based photometry and imaging, this confirms TOI-3686.01 as a \textit{bona fide} planet with a mass roughly half that of Jupiter.

However, the RV residuals show significant scatter, and their Lomb-Scargle periodogram \citep{LSPeriodogram_Lomb1976,LSPeriodogram_Scargle1982} peaks close to the stellar rotation period of $\Prot \sim 5.55$~days, motivating us to include a Gaussian Process (GP) model to account for activity-induced RV variations in our later modeling (Section \ref{sec:photodynamical_analysis}).

\section{Stellar Characterization} \label{sec:stellar_char}

\begin{deluxetable}{lcc}
\tablehead{
    \colhead{Parameter} & \colhead{Value} & \colhead{Source}
}
\tablecaption{TOI-3686 Stellar Parameters \label{tab:stellar_params}}
\startdata
% parameter, value, units
TIC ID & 13533330 & TICv8 \\
Gaia DR3 ID & 465296935580218368 & Gaia DR3 \\
2MASS ID & 02273616+5957098 & 2MASS \\
WISE ID & J022736.14+595709.8 & WISE \\
\multicolumn{3}{l}{\textbf{Astrometric Measurements}} \\
R.A. (J2016.0) & 02:27:36.13 & Gaia DR3 \\
Dec (J2016.0) & +59:57:09.79 & Gaia DR3 \\
$\mu_{\alpha*}$ (mas yr$^{-1}$) & $-14.012 \pm 0.012$ & Gaia DR3 \\
$\mu_{\delta}$ (mas yr$^{-1}$) & $-3.428 \pm 0.013$ & Gaia DR3 \\
$\varpi$ (mas) & $3.920 \pm 0.013$ & Gaia DR3 \\
\multicolumn{3}{l}{\textbf{Catalog Photometry}} \\
$T$ (mag) & $11.2380 \pm 0.0060$ & TICv8 \\
$G$ (mag) & $11.7659 \pm 0.0028$ & Gaia DR3 \\
$G_{BP}$ (mag) & $12.1898 \pm 0.0031$ & Gaia DR3 \\
$G_{RP}$ (mag) & $11.1681 \pm 0.0039$ & Gaia DR3 \\
$J$ (mag) & $10.427 \pm 0.020$ & 2MASS \\
$H$ (mag) & $10.148 \pm 0.018$ & 2MASS \\
$K_s$ (mag) & $10.038 \pm 0.016$ & 2MASS \\
W1 (mag) & $9.943 \pm 0.041$ & WISE \\
W2 (mag) & $9.988 \pm 0.020$ & WISE \\
W3 (mag) & $9.754 \pm 0.061$ & WISE \\
\multicolumn{3}{l}{\textbf{Spectroscopic Parameters}} \\
$T_{\rm eff}$ (K) & $6047 \pm 75$ & \texttt{SPC} \\
$\log{g}$ (cgs) & $4.35 \pm 0.13$ & \texttt{SPC} \\
$\feh$ (dex) & $0.086 \pm 0.080$ & \texttt{SPC} \\
$v\sin{i_\star}$ (\kms) & $11.64 \pm 0.50$ & \texttt{SPC} \\
\multicolumn{3}{l}{\textbf{Physical Stellar Parameters}} \\
$M_\star$ (\Msun) & $1.049^{+0.092}_{-0.085}$ & \Exofast \\
$R_\star$ (\Rsun) & $1.184^{+0.047}_{-0.044}$ & \Exofast \\
$T_{\rm eff}$ (K) & $5990^{+280}_{-270}$ & \Exofast \\
$\log{g}$ (cgs) & $4.313^{+0.050}_{-0.052}$ & \Exofast \\
$\feh$ (dex) & $0.042^{+0.084}_{-0.076}$ & \Exofast \\
Age (Gyr) & $5.9^{+4.4}_{-3.5}$ & \Exofast \\
Rot. Age (Gyr) & $0.73^{+0.62}_{-0.23}$ & \texttt{gyro-interp} \\
Li. Age (Gyr) & $1.27^{+2.2}_{-1.1}$ & \texttt{EAGLES}%

\enddata
\tablerefs{Catalog astrometry and photometry are drawn from \Gaia DR3 \citep{GaiaDR3_Vallenari2022}, 2MASS \citep{TMASS_Cutri2003,TMASS_Skrutskie2006}, WISE \citep{WISE_Cutri2012}, and the TESS Input Catalog \citep{TIC_Stassun2018,TIC_Stassun2019}.}
\end{deluxetable}

We used our high-resolution spectroscopic observations of TOI-3686 to measure stellar atmospheric parameters.
We applied the Stellar Parameter Classification code (SPC;
\citealt{SPC_Buchhave2012}) to match the observed TRES spectra against the \citet{Kurucz1993} grid of synthetic model spectra.
We took the mean and standard deviation as the final value and uncertainty of each parameter, adopting error floors of 50\,K in \Teff,
0.1~dex in \logg, 0.08~dex in \feh, and 0.5~\kms in \vsini to account for systematic model uncertainties.
This analysis yields $\Teff = 6047 \pm 75$~K, $\logg = 4.35 \pm 0.13$~dex, $\feh = 0.09\pm0.08$~dex, and $\vsini = 11.6 \pm 0.5$~\kms.

We also analyzed a co-added NEID spectrum with \texttt{SpecMatch-Empirical} \citep{SpecMatchEmp_Yee2017}, which compares the target spectrum to an empirical library of observed spectra of well-characterized stars.
We obtained stellar parameters consistent with the results from SPC: $\Teff = 6050 \pm 110$~K, $\Rstar = 1.12\pm0.20\,\Rsun$, and $\feh=0.04\pm0.09$~dex.
We adopt the SPC parameters, which have smaller uncertainties.

To complement the stellar atmospheric parameters, we used the \Exofast code \citep{ExoFAST_Eastman2013,ExoFASTv2_Eastman19} to constrain the physical parameters of TOI-3686 by modeling its broadband photometry with MIST stellar evolutionary models \citep{MIST0_Dotter2016,MISTI_Choi2016}.
We fitted photometry from the \Gaia DR3, 2MASS, and WISE catalogs, imposing error floors of 0.02, 0.02, and 0.03~mag on the photometric uncertainties respectively.
We also used as input the zero-point corrected parallax from \Gaia \citep{GaiaDR3_Vallenari2022,GaiaEDR3_Lindegren2021}.
We imposed a Gaussian prior on \feh from our \texttt{SPC} analysis and an upper limit on the visual extinction $A_V$ from galactic dust maps from \citet{Schlegel1998,Schlafly2011}.
We sampled the posterior using a differential
evolution Markov Chain Monte Carlo (DE-MCMC) algorithm, requiring a Gelman-Rubin statistic below 1.01 for convergence \citep{GelmanRubin}.

Our final stellar properties are presented in Table \ref{tab:stellar_params}, indicating that TOI-3686 is a star slightly more massive and metal-rich than the Sun.
The MIST modeling provides only a wide constraint on the age of TOI-3686; however, a few indicators of youth suggest that it may have an age $\lesssim 1$~Gyr, which we discuss below.

\subsection{Indications of youth for TOI-3686}
\label{ssec:youth}

The TESS light curves of TOI-3686 show significant, $1$--2\% peak-to-peak quasi-periodic variability that is likely due to rotational modulation.
Modeling this with a GP (\S\ref{ssec:transit_search}), we measure a mean $\Prot = 5.55 \pm 0.06$~days.
Assuming that its rotation has only been modified by magnetic braking, this moderately rapid rotation indicates that TOI-3686 is somewhat youthful compared to the typical field star.
We used the \texttt{gyro-interp} code \citep{gyrointerp_Bouma2023} to infer a gyrochronological age of $730^{+620}_{-230}$~Myr.
One caveat is that gyrochronological ages may not be reliable for stars that host hot Jupiters due to tidal spin-up \citep[e.g.,][]{Brown2014,Maxted2015,TejadaArevalo2021}, although the relatively low mass ($\Mp\approx0.5\,\Mjup$) and long orbital period ($P = 5.078$~days) of TOI-3686.01 would weigh against significant modification of its host star's spin period.

Another indicator of youth is the presence of lithium in the stellar atmosphere, as stars with significant convective envelopes should transport Li into the interior where it is destroyed, depleting photospheric Li over time.
Using the high S/N stacked NEID spectrum (\S\ref{sec:stellar_char}), we detected significant absorption in the Li I 6707.8\,\AA~line, with an equivalent width (EW) of $51.8^{+1.6}_{-1.9}\,\mathrm{m\AA}$.
This was measured following the procedure from \citet{Bouma2024}, which corrects for blending of the Li line with a nearby Fe line by subtracting a calibrated mean offset.
We then used the \texttt{EAGLES} code \citep{Jeffries2023} to estimate the stellar age given the stellar \Teff and Li EW measurement, finding an age of $1270^{+2200}_{-1100}$~Myr.
The weaker constraining power of the Li method for stars with this \Teff is expected, given the limited depth of their convection zones.
Still, both the Li absorption and gyrochronology appear to point consistently to a relatively young, $\lesssim 1$~Gyr age for TOI-3686.

\section{Light Curve Analysis \label{sec:lc_analysis}}
\subsection{Transit Time Extraction \label{ssec:transit_times}}

To characterize the TTVs of TOI-3686.01, we measured individual transit times from the space- and ground-based observations.
We used the \exoplanet code \citep{exoplanet_ForemanMackey2021} to model the photometry, which implements the differentiable transit model of \citet{exoplanet_Agol2020} for rapid evaluation of transit light curves and their derivatives.
We modeled each observation with its own mid-transit time $T_{c,i}$ and derived a mean orbital period \Porb from a least-squares fit to these transit times.
The other transit parameters $b$, \ror, and stellar pseudo-density \rhostarc, were shared across all transits, and with broad uniform priors on these parameters.
We assumed a quadratic limb-darkening law, with coefficients shared across observations in the same filter, sampling in the $(q_1,\,q_2)$ basis from \citet{limbdarkening_Kipping2013}.

We simultaneously detrended each ground-baesd light curve against a linear combination of instrumental systematics such as pixel position, airmass, and point spread function (PSF) width.
For the TESS observations, we first masked the transit events and fitted a basis spline to the out-of-transit flux with the \texttt{Keplerspline} code to remove the long-term variability \citep{Keplerspline_Vanderburg2014,Keplerspline_Shallue2018}.
We then split each sector into separate light curves for each transit, retaining $\pm 2$ transit durations of data around the transit center.
To account for remaining uncorrected flux variations in the TESS data, we allowed for simultaneous detrending against a quadratic function of time.

We report the best-fit transit times and 1-$\sigma$ uncertainties in Table \ref{tab:transit_times}.
The TTVs appear well-described by a sinusoidal model, with a peak-to-peak amplitude of 40~minutes, some of the largest ever seen for a hot Jupiter, and a period of $\sim 260$~days (Figure \ref{fig:photodynamical_results}).
These TTVs clearly indicate the presence of a nearby planet in or close to a mean-motion resonance (MMR; e.g. \citealp{Lithwick2012b,Nesvorny2016}), motivating us to search for additional transits in the \TESS data.

\subsection{Detection of a Second Transiting Planet}
\label{ssec:transit_search}

\begin{figure*}
    \centering
    \includegraphics[width=\linewidth]{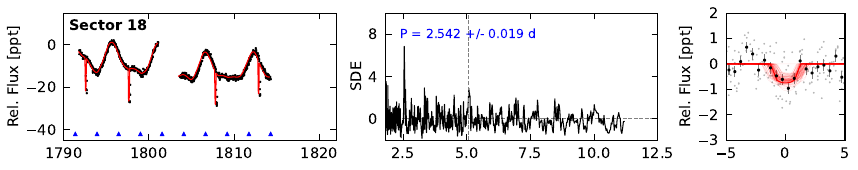}
    \includegraphics[width=\linewidth]{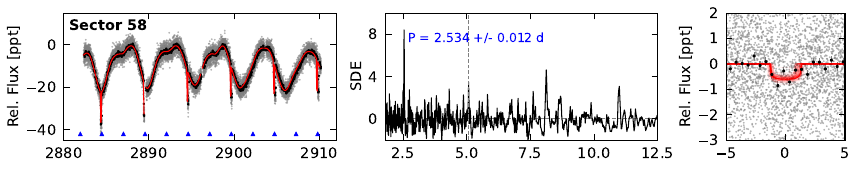}
    \includegraphics[width=\linewidth]{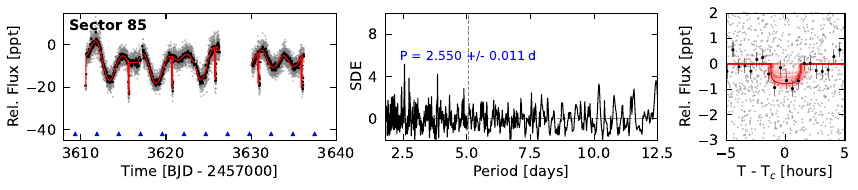}
    \caption{\textbf{Left:} TESS photometry of TOI-3686 from Sectors 18, 58, and 85 (top to bottom), at native cadence (light gray) and binned to 1800s cadence (black).
    The maximum likelihood two-planet and GP model for the photometry is shown in red.
    The hot Jupiter's transits and stellar rotational modulation are clearly visible; the transits of the small inner planet are not visible on this scale, but the times at which they occur are marked with blue triangles.
    \textbf{Middle:} TLS periodogram for each sector after subtracting a one-planet and GP model, with the highest peak labeled.
    Vertical dashed lines mark the hot Jupiter's orbital period.
    \textbf{Right:} The same residual photometry phase-folded on the inner planet's period, with 100 posterior draws of its transit model in red.
    }
    \label{fig:tess_periodograms}
\end{figure*}

Previous searches of the \TESS data had not detected any other transits, but shallow events could have been missed given the stellar variability and potentially large TTVs.
We thus decided to perform our own search of the \TESS light curves.

We again used the \exoplanet code \citep{exoplanet_ForemanMackey2021} to model the TESS data, this time using the SPOC Simple Aperture Photometry (SAP) light curves, as these retained more of the stellar variability signal than those that have undergone Pre-search Data Conditioning (PDCSAP light curves).
We fitted the full light curve from each sector separately to allow for changes in stellar variability and the orbital ephemeris due to TTVs.
We first included only the hot Jupiter's transits and a GP model for the stellar variability, where we used the \texttt{RotationTerm} GP kernel implemented in \texttt{celerite2} \citep{celerite1_ForemanMackey2017,celerite2_ForemanMackey2018}, which models stellar rotation as a mixture of two stochastically-driven and damped harmonic oscillators.
We assumed a constant linear ephemeris within each sector, since we found no evidence for high-frequency ``chopping'' signals, while the $\sim25$~day sector baseline is only 10\% of the TTV super-period such that the long-term TTV signal would be absorbed by changes in the mean orbital period and time of conjunction.
We then optimized over the hot Jupiter's transit and orbital parameters and the GP hyperparameters.

We then subtracted the MAP single-planet and GP model from the data, and searched the residuals for additional transit signals using the transit least squares (TLS) algorithm \citep{tls_Hippke2019}.
A peak at $\sim 2.54$ days, roughly half the hot Jupiter's orbital period, is the most significant periodogram peak in all three sectors, with Signal Detection Efficiencies (SDEs) of 6.1, 7.4, and 5.3 respectively (Figure \ref{fig:tess_periodograms}).
No statistically significant peaks were seen at either the hot Jupiter's orbital period or the stellar rotation period, indicating that our model removed those signals effectively.

We then added the transits of a second planet to our model, placing uniform priors on its orbital period and time of transit centered on the periodogram peaks.
As before, we assumed a constant period within each sector. 
We used the Hamiltonian Monte Carlo (HMC) No U-Turn Sampler \citep{NUTS_Hoffman} implemented in \texttt{pymc} \citep{PyMC} to explore the posterior probability distributions of the fitted parameters in this model.
Each sector's fit was conducted independently, yet we found properties for the second planet that were consistent between them; notably the transit depths yielded $R_p/R_\star = 0.024\pm0.006,\,0.023\pm0.003$, and $0.024\pm0.007$ respectively, strongly suggesting that they are indeed due to a planet and not spurious noise.
The transit ephemerides of this second planet vary between sectors, but we show in Section \ref{sec:photodynamical_analysis} that these variations are consistent with a TTV signal $180^\circ$ out of phase with that of the hot Jupiter.
The best-fit transit models along with random draws from the posterior, are plotted in Figure \ref{fig:tess_periodograms}.

Given that the transit depths are measured at 4-, 7- and 3.5-$\sigma$ significance in individual sectors with consistent shapes and depths, we regard this as a confident detection of a second transiting planet in the TOI-3686 system. 
The transits correspond to a sub-Neptune-sized ($\Rp \approx 3\,\Rearth$) planet in a 1:2 period commensurability with the hot Jupiter.
In the remainder of this paper, we focus our analysis on this system architecture, referring to the inner sub-Neptune as TOI-3686\,b and the outer hot Jupiter as TOI-3686\,c.

\begin{rotatepage}
\movetabledown=0.5in
\begin{longrotatetable}
\begin{deluxetable*}{cccccccc}
\tabletypesize{\footnotesize}
\tablecolumns{8}
\tablecaption{Photometric Observations and Mid-Transit Times of TOI-3686\,c
\label{tab:transit_times}}
\tablehead{
    \colhead{Epoch} &
    \colhead{Facility/Instrument} & \colhead{Location} & \colhead{Aperture} & \colhead{Filter} &
    \colhead{Date} & \colhead{Cadence} & \colhead{Transit Time\tablenotemark{a}} \\
    & & & \colhead{(m)} & & \colhead{(UT)} & \colhead{(s)} & \colhead{(\bjdtdb)}
}
\startdata
% epoch, observatory, location, size, filter, ut, cadence, transit_time
0 & TESS Sector 18 & Space & 0.1 & TESS & 2019 Nov 05 & 1800 & $2458792.55799(67)(78)$ \\
1 & TESS Sector 18 & Space & 0.1 & TESS & 2019 Nov 10 & 1800 & $2458797.63638(71)(68)$ \\
3 & TESS Sector 18 & Space & 0.1 & TESS & 2019 Nov 20 & 1800 & $2458807.78781(70)(66)$ \\
4 & TESS Sector 18 & Space & 0.1 & TESS & 2019 Nov 25 & 1800 & $2458812.86795(84)(59)$ \\
138 & OAUV/T50 & Valencia, Spain & 0.5 & $R$ & 2021 Oct 05 & 75 & $2459493.35915(94)(90)$ \\
143 & Waffelow Creek Observatory & Nacogdoches, TX, USA & 0.36 & $r^\prime$ & 2021 Oct 31 & 60 & $2459518.74545(59)(48)$ \\
152 & Observatory de Ca l'Ou & Sant Martí Sesgueioles, Barcelona, Spain & 0.4 & $R$ & 2021 Dec 15 & 100 & $2459564.4406(10)(11)$ \\
153 & SUTO/OTIVAR & Pyskowice, Poland & 0.3 & $R$ & 2021 Dec 20 & 90 & $2459569.51586(91)(81)$ \\
215 & TESS Sector 58 & Space & 0.1 & TESS & 2022 Oct 31 & 120 & $2459884.34713(61)(50)$ \\
216 & TESS Sector 58 & Space & 0.1 & TESS & 2022 Nov 05 & 120 & $2459889.42431(82)(69)$ \\
217 & TESS Sector 58 & Space & 0.1 & TESS & 2022 Nov 11 & 120 & $2459894.50254(55)(45)$ \\
218 & TESS Sector 58 & Space & 0.1 & TESS & 2022 Nov 16 & 120 & $2459899.58002(53)(52)$ \\
219 & TESS Sector 58 & Space & 0.1 & TESS & 2022 Nov 21 & 120 & $2459904.65817(52)(51)$ \\
219 & Acton Sky Portal & Acton, MA, USA & 0.36 & $g^\prime$ & 2022 Nov 21 & 32 & $2459904.65646(40)$ \\
220 & TESS Sector 58 & Space & 0.1 & TESS & 2022 Nov 26 & 120 & $2459909.73720(51)(57)$ \\
278 & Kourovka Observatory & Yekaterinburg, Russia & 0.4 & $R$ & 2023 Sep 16 & 240 & $2460204.26530(88)(92)$ \\
278 & Kourovka Observatory & Yekaterinburg, Russia & 0.4 & $V$ & 2023 Sep 16 & 240 & $2460204.2694(13)$ \\
337 & Pine Mountain Observatory/CDK14 & Bend, OR, USA & 0.35 & $r^\prime$ & 2024 Jul 12 & 110 & $2460503.88499(49)(56)$ \\
344 & OAUV/TURIA2 & Valencia, Spain & 0.3 & $R$ & 2024 Aug 16 & 195 & $2460539.44256(51)$ \\
345 & Kotizarovci Observatory & Viskovo, Croatia & 0.3 & $V$ & 2024 Aug 21 & 120 & $2460544.51810(98)(72)$ \\
348 & Peters Observatory & Clinton, NY, USA & 0.5 & $r^\prime$ & 2024 Sep 05 & 60 & $2460559.75519(25)(28)$ \\
348 & Nube Blanca Observatory & Berthoud, CO, USA & 0.4 & $R$ & 2024 Sep 06 & 160 & $2460559.75356(72)(81)$ \\
349 & FLWO/KeplerCam & Amado, AZ, USA & 1.2 & $i^\prime$ & 2024 Sep 11 & 40 & $2460564.83296(18)(17)$ \\
349 & LCO McDonald/Sinistro & Fort Davis, TX, USA & 1.0 & $z^\prime$ & 2024 Sep 11 & 67 & $2460564.83224(23)(24)$ \\
349 & Lookout Observatory & Colorado Springs, CO, USA & 0.28 & $V$ & 2024 Sep 11 & 20 & $2460564.8311(14)(12)$ \\
349 & Union College Observatory & Schenectady, NY, USA & 0.5 & $V$ & 2024 Sep 11 & 60 & $2460564.83450(25)$ \\
358 & Acton Sky Portal & Acton, MA, USA & 0.36 & $r^\prime$ & 2024 Oct 26 & 20 & $2460610.52944(30)(31)$ \\
358 & Privat Observatory Herges-Hallenberg & Steinbach-Hallenberg, Germany & 0.28 & $BB$ & 2024 Oct 26 & 120 & $2460610.5280(37)(30)$ \\
359 & TESS Sector 85 & Space & 0.1 & TESS & 2024 Nov 01 & 120 & $2460615.60747(57)$ \\
360 & TESS Sector 85 & Space & 0.1 & TESS & 2024 Nov 06 & 120 & $2460620.68311(59)(54)$ \\
360 & Mason Observatory & Oklahoma City, OK, USA & 0.28 & $V$ & 2024 Nov 05 & 90 & $2460620.6848(11)(13)$ \\
360 & LCO McDonald/Sinistro & Fort Davis, TX, USA & 1.0 & $z^\prime$ & 2024 Nov 06 & 95 & $2460620.68322(24)(26)$ \\
361 & LCO McDonald/Sinistro & Fort Davis, TX, USA & 1.0 & $z^\prime$ & 2024 Nov 11 & 67 & $2460625.75964(23)(25)$ \\
362 & TESS Sector 85 & Space & 0.1 & TESS & 2024 Nov 16 & 120 & $2460630.83783(57)(61)$ \\
362 & LCO McDonald/Sinistro & Fort Davis, TX, USA & 1.0 & $z^\prime$ & 2024 Nov 16 & 67 & $2460630.83758(47)(43)$ \\
363 & TESS Sector 85 & Space & 0.1 & TESS & 2024 Nov 21 & 120 & $2460635.91428(51)(49)$ \\
369 & Kotizarovci Observatory & Viskovo, Croatia & 0.3 & $V$ & 2024 Dec 21 & 120 & $2460666.3719(16)(17)$ \\
371 & LCO Teide/Sinistro & La Laguna, Tenerife, Spain & 1.0 & $i^\prime$ & 2024 Dec 31 & 17 & $2460676.52554(32)(28)$ \\
373 & FLWO/KeplerCam & Amado, AZ, USA & 1.2 & $i^\prime$ & 2025 Jan 11 & 60 & $2460686.68456(24)(29)$ \\
374 & BYU OPO & Provo, UT, USA & 0.3 & $V$ & 2025 Jan 16 & 75 & $2460691.7617(18)(24)$ \\
411 & OAUV/TURIA2 & Valencia, Spain & 0.3 & $R$ & 2025 Jul 22 & 195 & $2460879.66949(77)(72)$ \\
421 & OAUV/TURIA2 & Valencia, Spain & 0.3 & $R$ & 2025 Sep 11 & 195 & $2460930.43795(43)(50)$ \\
422 & OAUV/TURIA2 & Valencia, Spain & 0.3 & $R$ & 2025 Sep 16 & 195 & $2460935.51959(40)(36)$ \\
423 & LCO Teide/Sinistro & La Laguna, Tenerife, Spain & 1.0 & $i^\prime$ & 2025 Sep 21 & 42 & $2460940.59470(20)(19)$ \\
424 & LCO Teide/Sinistro & La Laguna, Tenerife, Spain & 1.0 & $i^\prime$ & 2025 Sep 27 & 42 & $2460945.67077(20)(18)$ \\
424 & Union College Observatory & Schenectady, NY, USA & 0.5 & $R$ & 2025 Sep 27 & 250 & $2460945.6704(13)(12)$ \\
425 & Pritchard Observatory & York, PA, USA & 0.2 & $V$ & 2025 Oct 01 & 60 & $2460950.74219(46)(39)$ \\
425 & Union College Observatory & Schenectady, NY, USA & 0.5 & $R$ & 2025 Oct 02 & 69 & $2460950.7495(13)(15)$ \\
426 & BYU OPO & Provo, UT, USA & 0.6 & $R$ & 2025 Oct 07 & 150 & $2460955.82511(48)(46)$ \\
433 & Maidanak Astronomical Observatory & Qashqadaryo, Uzbekistan & 0.6 & $R$ & 2025 Nov 11 & 60 & $2460991.37734(66)(56)$ \\
437 & LCO McDonald/Sinistro & Fort Davis, TX, USA & 1.0 & $i^\prime$ & 2025 Dec 02 & 17 & $2461011.69166(20)(21)$ \\
437 & Nube Blanca Observatory & Berthoud, CO, USA & 0.4 & $R$ & 2025 Dec 02 & 60 & $2461011.6902(12)(11)$ \\
439 & Unistellar evScope & Placitas, NM, USA & 0.1 & $V$ & 2025 Dec 12 & 120 & -- \\
445 & Privat Observatory Herges-Hallenberg & Steinbach-Hallenberg, Germany & 0.28 & $V$ & 2026 Jan 11 & 120 & $2461052.3283(11)(09)$ \\
446 & Salerno University Observatory & Salerno, Italy & 0.6 & $R$ & 2026 Jan 16 & 120 & $2461057.4076(10)(11)$ \\
448 & Marvel Telescope & La Palma, Spain & 0.8 & $g^\prime$ & 2026 Jan 26 & 45 & $2461067.56733(49)(58)$ \\
486 & SUTO/OTIVAR & Pyskowice, Poland & 0.3 & $R$ & 2026 Aug 07 & 90 & $2461260.5154(16)(15)$ \\
488 & Two-meter Twin Telescope & La Laguna, Tenerife, Spain & 0.8 & $i^\prime$ & 2026 Aug 18 & 40 & $2461270.67621(61)(78)$ \\
489 & Two-meter Twin Telescope & La Laguna, Tenerife, Spain & 0.8 & $i^\prime$ & 2026 Aug 23 & 40 & $2461275.7549(15)$%
\enddata
\tablenotetext{a}{The numbers in parentheses represent the uncertainty in the last two decimal digits. If two sets of numbers are provided, these represent the upper and lower 1-$\sigma$ uncertainties respectively.}
\tablecomments{Acronyms for facility names are provided here. OAUV: Observatori Astronòmic de la Universitat de València; SUTO: Silesian University of Technology Observatories; FLWO: Fred Lawrence Whipple Observatory; LCO: Las Cumbres Observatory \citep{LCOGT_Brown2013}; BYU OPO: Brigham Young University Orson Pratt Observatory.}
\end{deluxetable*}
\end{longrotatetable}
\end{rotatepage}

\section{Photodynamical Analysis}
\label{sec:photodynamical_analysis}

\begin{figure*}
    \centering
    \includegraphics[width=\linewidth]{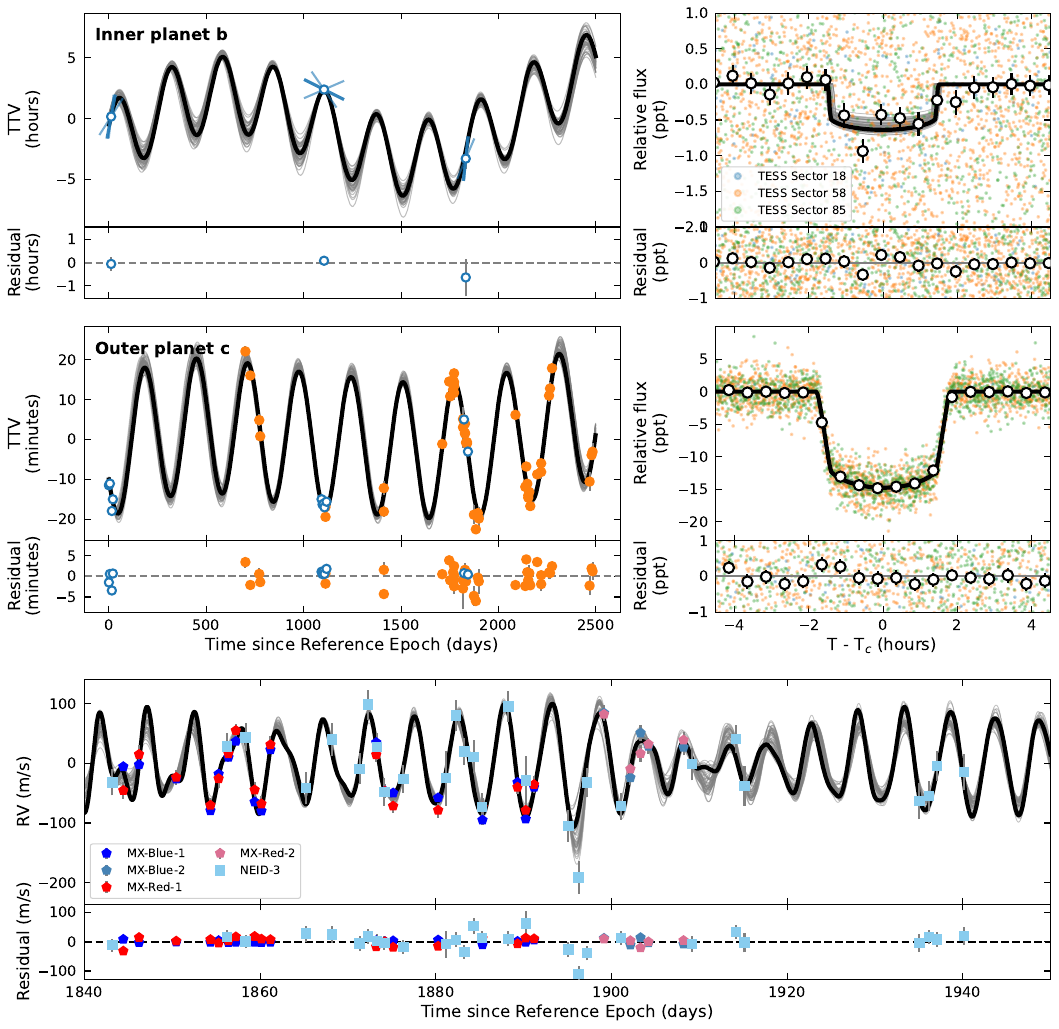}
    \caption{Results of our photodynamical modeling. In all panels, the MAP model is shown as the thick black line, while 50 random posterior realizations are plotted in gray.
    \textbf{Top two rows, left:} TTVs of planets b and c (note the different time units). Filled orange circles correspond to measured ground-based transit times. The open blue circles show TESS transit times, shown for reference only since the TESS photometry is modeled directly. For planet b, the blue points represent the osculating time of conjunction in each sector, while the tangent lines shows the MAP value of the osculating period (dark blue) and 1-$\sigma$ posterior range (light blue), as measured in \S\ref{ssec:transit_search}.
    \textbf{Top two rows, right:} TESS photometry from all three sectors, detrended by the MAP GP model and with each transit phased to its individual predicted transit time, i.e. with TTVs removed. Open circles show the data binned to 1800s cadence.
    \textbf{Bottom:} Stellar RVs, showing reflex motion due to the planets and rotation-modulated activity modeled by a GP. A per-instrument offset has been removed, and error bars include the instrumental uncertainties and a per-instrument jitter term added in quadrature. (The RV measurements used to create this figure are available as a machine-readable table accompanying the online article.)
    \digitalasset
    \label{fig:photodynamical_results}}
\end{figure*}

To leverage the rich information encoded by the TTVs about the planetary masses, eccentricities and dynamical states, we modeled the TOI-3686 planets with $3$-body integrations.
We used the \texttt{jnkepler} code \citep{jnkepler_Masuda2024,jnkepler_ASCL_Masuda2025}, an $N$-body code written in \texttt{JAX}, which enables automatic differentiation of the model with respect to the fitted parameters and efficient sampling of their posterior distributions with HMC algorithms.
Given a set of initial conditions for the two planets, we integrate the evolution of the system for 2500~days, covering full observational baseline.

Because the transits of the inner sub-Neptune are too shallow to measure individual transit times, we implemented a ``photodynamical'' model, directly fitting the TESS flux time series.
In \texttt{jnkepler}, the flux loss due to each planet's transits is computed using the modeled sky-plane positions and velocities at the transit mid-point, with the \citet{exoplanet_Agol2020} transit model implemented in \texttt{jaxoplanet} \citep{jaxoplanet_Hattori2024}.
The model for Sector 18 was evaluated at 2-minute cadence and integrated over the actual 30-minute observation cadence to account for the finite integration time \citep{binning_Kipping2010}.
We also allowed for potential dilution of the transit depth from nearby stars by fitting for a flux dilution term $D_s$ \citep[e.g.,][]{Han2025}, as well as a flux jitter term to account for underestimated flux uncertainties.
As in Section \ref{ssec:transit_search}, we modeled the stellar variability in each TESS sector with a \texttt{RotationTerm} GP kernel.
Thus, the likelihood for the data from TESS sector $s$ is given by:
\begin{align}
\begin{split}
    \ln\mathcal{L}_{\mathrm{photodyn},s} = \ln\boldsymbol{\mathcal{N}}(\mathbf{r}_s\, |\, \mathbf{0}\,,\, &\mathbf{K}_{\mathrm{rot},s} +  \\ &\mathrm{diag}\!\left(\sigma_{\mathrm{obs},s}^2 + \sigma_{\mathrm{jit},s}^2\right)),
\label{eq:photodyn_likelihood}
\end{split}
\end{align}
where $\boldsymbol{\mathcal{N}}(\mathbf{r}_s\,|\,\boldsymbol{\mu},\,\boldsymbol{\sigma}^2)$ denotes the multivariate Gaussian probability density evaluated at $\mathbf{r}_s$; $\mathbf{K}_{\mathrm{rot},s}$ is the \texttt{RotationTerm} kernel; and $\mathbf{r}_s$ is the residual vector taken by subtracting the 2-planet transit model from the observed fluxes.

To save on computational cost, we did not model the ground-based photometry directly but fitted the mid-transit times for those observations extracted in Section \ref{ssec:transit_times}.
We assumed a standard Gaussian likelihood, including a global timing jitter term added in quadrature with each timing measurement's uncertainty.

Finally, our preliminary analysis (\S\ref{sssec:rv_analysis}) revealed that the RV data also contain a significant activity-induced signal with periodic modulation at the stellar rotation period.
Thus, our model for the RV data includes both the stellar velocities computed from the $N$-body integration at the actual observation times, as well as a GP model for stellar variability, again using the \texttt{RotationTerm} kernel.
Because the majority of the RV data are contemporaneous with the TESS Sector 85 observations, we shared the rotation period hyperparameter $P_\mathrm{GP}$ between the RV GP model and that for the Sector 85 photometry, but treated the remaining GP hyperparameters independently.
We also fitted for a per-instrument offset $\gamma_i$ and jitter, such that each instrument $i$ contributes the likelihood:
\begin{align}
\begin{split}
    \ln\mathcal{L}_{\mathrm{RV},i} = \ln\boldsymbol{\mathcal{N}}(\mathbf{r}_i\, |\, \mathbf{0}\,,\, &\mathbf{K}_{\mathrm{rot,RV}} +  \\ &\mathrm{diag}\!\left(\sigma_{\mathrm{obs},i}^2 + \sigma_{\mathrm{jit},i}^2\right)),
\label{eq:rv_likelihood}
\end{split}
\end{align}
and $\mathbf{r}_i = \mathbf{RV}_{\mathrm{obs},i} - \mathbf{RV}_\mathrm{N\text{-}body} - \gamma_i$ is the residual vector for the RV observations.

The total log-likelihood is then
\begin{align}
\begin{split}
    \ln\mathcal{L_\mathrm{tot}} = &\sum_{s}\ln\mathcal{L}_{\mathrm{photodyn},s} +
    \ln\mathcal{L}_\mathrm{TTV} + \sum_i\ln\mathcal{L}_{\mathrm{RV},i},
\label{eq:total_likelihood}
\end{split}
\end{align}
where the first sum is over the three TESS sectors and the second is over the five RV instruments.

We varied all orbital elements, masses, and $\Rp/\Rstar$ for both planets. 
Instead of mean anomaly at the initial epoch, we sampled the osculating time of conjunction at the start of integration, which allowed more intuitive bounds to be placed on this parameter.
We parameterized the planetary inclinations in terms of impact parameters and bounded them between 0 and 1.
To convert from mass and radius ratios with the star to physical planet properties, we also fitted for the stellar mass and radius, imposing Gaussian priors centered at the values from Section \ref{sec:stellar_char}.
We note that the stellar mass and radius are linked in the photodynamical model because the transit durations depend on the stellar density for a given orbital period.
Because the fitted dilution terms are degenerate with the planet-to-star radius ratios in affecting the measured TESS transit depths, we imposed a Gaussian prior on $\Rp/\Rstar$ for the hot Jupiter, derived from a fit similar to that described in \S\ref{ssec:transit_times}, but using the ground-based data only. Since the ground-based observations resolved TOI-3686 from all neighboring bright stars, we treat the transit depth derived from those observations as the undiluted ground truth.
The remaining fitted parameters received broad uniform or log-uniform priors, listed in Table \ref{tab:photodynamical_fit}.

We first investigated the solution landscape by conducting a grid search of possible initial conditions: we randomly sampled system parameters within their prior ranges, then performed a local optimization over the model posterior probability with the \texttt{L-BFGS-B} algorithm.
We found multiple modes corresponding to different orbital periods of TOI-3686\,b, arising from a different integer number of elapsed orbits between TESS sectors.
We then performed a fine sweep in $P_b$ to more closely sample each alias.
Ultimately, we found that the mode corresponding to 722 orbits of TOI-3686\,b between the reference $T_0$ in Sectors 18 and 85 was most favored, with the next best mode having significantly lower posterior probability ($\Delta\ln{p} = -33.8$). 

We used the maximum a posteriori (MAP) result from this mode to initialize a \texttt{numpyro} HMC sampler \citep{numpyro_Bingham2019,numpyro_Phan2019}, running four parallel HMC chains with 1500 warm-up steps and 1000 sampling steps.
We checked for convergence by ensuring the rank-normalized $\widehat{R}$ statistic \citep{Rhat_Vehtari2021} was below 1.01 for all parameters, and an effective sample size $\text{ESS}> 500$.

Figure \ref{fig:photodynamical_results} shows the data and predictions from the best-fitting photodynamical model, while the parameter inferences are reported in Table \ref{tab:photodynamical_fit}.
The GP hyperparameters for the stellar flux variability model are similar to those obtained in our earlier light curve analysis (\S\ref{ssec:transit_search}), while for the RVs, the activity GP model has a standard deviation $\sigma = 30~\ms$.
The predicted TTV model for TOI-3686\,c is essentially sinusoidal, with a peak-to-peak amplitude of 40~minutes, while the TTVs of the inner planet have a peak-to-peak amplitude of 10 hours, $\sim 16\%$ of its orbital period.
The period ratio oscillates around a mean value of $P_c/P_b = 2.0025$, corresponding to a deviation from perfect period commensurability of  
\begin{equation}
\Delta \equiv \frac{P_\mathrm{c}}{P_\mathrm{b}}\frac{j-1}{j} - 1 = 0.00124.    
\label{eq:Delta}
\end{equation}
This small value of $\Delta$ indicates that the system is likely in, not merely close to, an MMR \citep{Goldberg2023}, which we explore further in Section \ref{sec:dynamics}.

The modeling of the RV and TTV data allow us to constrain the masses of both planets.
We find that TOI-3686\,c has a mass of $\Mp = 0.56\pm0.04\,\Mjup$, consistent with that measured in our preliminary RV fit (\S\ref{sssec:rv_analysis}).
With an equilibrium temperature of $\Teq = 1300\,\mathrm{K}$, the planet is subject to the ``hot Jupiter radius anomaly'' that is active above $\Teq \gtrsim 1000\,\mathrm{K}$ \citep{Demory2011,Miller2011}.
The measured $\Rp = 1.27\pm0.03\,\Rjup$ is consistent with empirical relationships for this anomalous inflation \citep{Thorngren2021} and is significantly inflated from the $\Rp \sim 0.9\,\Rjup$ that would be expected for a planet of this mass at lower insolation levels \citep{Thorngren2019}.
Meanwhile, TOI-3686\,b has a mass and radius typical of a sub-Neptune, ($\Mp = 10.5^{+1.3}_{-1.2}\,\Mearth$, $\Rp = 2.8\pm0.2\,\Rearth$).
We note that this mass constraint comes entirely from the TTVs of TOI-3686\,c, as its expected RV semi-amplitude of $K \approx 5\,\ms$ is too small to be detected by the RV data alone.

\startlongtable
\begin{deluxetable*}{lcc}
\tablecolumns{3}
\tablecaption{Photodynamical Modeling Results \label{tab:photodynamical_fit}}
\tablehead{
    \colhead{Parameter} & \colhead{Prior} & \colhead{Posterior}
}
\startdata%
% parameter, prior, value
\\[-\normalbaselineskip]\multicolumn{3}{l}{\textbf{Planet b}} \\
$P_b$ (days) & $\mathcal{U}(2.4886, 2.5389)$ & $2.54320^{+0.00066}_{-0.00086}$ \\
$T_{c,b}$ ($\mathrm{BJD} - T_\mathrm{ref}$) & $\mathcal{U}(0.8, 1.7)$ & $1.384^{+0.015}_{-0.013}$ \\
$e_b\cos{\omega_b}$ & $\mathcal{U}(-0.4, 0.4)$\tablenotemark{a} & $0.019^{+0.013}_{-0.009}$ \\
$e_b\sin{\omega_b}$ & $\mathcal{U}(-0.4, 0.4)$\tablenotemark{a} & $-0.1584 \pm 0.0026$ \\
$e_b$ & Derived\tablenotemark{a} & $0.1595^{+0.0013}_{-0.0016}$ \\
$M_b$ ($M_\odot$) & $\mathcal{LU}(10^{-6}, 10^{-3})$ & $0.0000315^{+0.0000040}_{-0.0000036}$ \\
$M_b$ ($M_\oplus$) & Derived & $10.5^{+1.3}_{-1.2}$ \\
$R_b/R_\star$ & $\mathcal{U}(0.001, 0.2)$ & $0.0231^{+0.0015}_{-0.0014}$ \\
$R_b$ ($R_\oplus$) & Derived & $2.84^{+0.19}_{-0.20}$ \\
$b_b$ & $\mathcal{U}(0, 1)$ & $0.41^{+0.18}_{-0.24}$ \\
$a_b/R_\star$ & Derived & $7.23 \pm 0.14$ \\
\multicolumn{3}{l}{\textbf{Planet c}} \\
$P_c$ (days) & $\mathcal{U}(5.0270, 5.1286)$ & $5.076790^{+0.000074}_{-0.000068}$ \\
$T_{c,c}$ ($\mathrm{BJD} - T_\mathrm{ref}$) & $\mathcal{U}(-0.4, 0.4)$ & $0.00053^{+0.00041}_{-0.00046}$ \\
$e_c\cos{\omega_c}$ & $\mathcal{U}(-0.4, 0.4)$\tablenotemark{a} & $0.00024^{+0.00084}_{-0.00074}$ \\
$e_c\sin{\omega_c}$ & $\mathcal{U}(-0.4, 0.4)$\tablenotemark{a} & $-0.0006^{+0.0010}_{-0.0011}$ \\
$e_c$ & Derived\tablenotemark{a} & $0.0012^{+0.0005}_{-0.0011}$ \\
$M_c$ ($M_\odot$) & $\mathcal{LU}(10^{-6}, 10^{-3})$ & $0.000538^{+0.000034}_{-0.000038}$ \\
$M_c$ ($M_\mathrm{Jup}$) & Derived & $0.564^{+0.036}_{-0.040}$ \\
$R_c/R_\star$ & $\mathcal{N}(0.1158, 0.0005)$ & $0.11575^{+0.00050}_{-0.00051}$ \\
$R_c$ ($R_\mathrm{Jup}$) & Derived & $1.272^{+0.038}_{-0.028}$ \\
$b_c$ & $\mathcal{U}(0, 1)$ & $0.287^{+0.066}_{-0.056}$ \\
$a_c/R_\star$ & Derived & $11.46^{+0.23}_{-0.22}$ \\
$\Delta\Omega_{cb}$ (deg) & $\mathcal{U}(-180, 180)$ & $-7.1^{+2.7}_{-2.9}$ \\
$i_\mathrm{mut}$ (deg) & Derived & $7.5^{+2.3}_{-2.8}$ \\
\multicolumn{3}{l}{\textbf{Stellar Parameters}} \\
$M_\star$ ($M_\odot$) & $\mathcal{N}(1.049, 0.09)$ & $1.130^{+0.075}_{-0.074}$ \\
$R_\star$ ($R_\odot$) & $\mathcal{N}(1.184, 0.045)$ & $1.130^{+0.030}_{-0.027}$ \\
$u_{1,\mathrm{TESS}}$ & $\mathcal{U}(0, 1)$\tablenotemark{b} & $0.358^{+0.088}_{-0.069}$ \\
$u_{2,\mathrm{TESS}}$ & $\mathcal{U}(0, 1)$\tablenotemark{b} & $-0.05^{+0.10}_{-0.17}$ \\
\multicolumn{3}{l}{\textbf{TTV Jitter}} \\
$\sigma_{\mathrm{TTV}}$ (days) & $\mathcal{LU}(10^{-5}, 10^{-1})$ & $0.00139^{+0.00017}_{-0.00022}$ \\
\multicolumn{3}{l}{\textbf{TESS Sector 18 Photometry}} \\
$F_{0,\mathrm{S18}}$ & $\mathcal{U}(-0.1, 0.1)$ & $0.00170^{+0.00064}_{-0.00059}$ \\
$\sigma_{\mathrm{GP,S18}}$ (ppt) & $\mathcal{LU}(0.1, 50)$ & $9.3^{+2.7}_{-4.3}$ \\
$P_{\mathrm{GP,S18}}$ (days) & $\mathcal{U}(4, 20)$ & $5.60^{+0.12}_{-0.15}$ \\
$\log{Q_{0,\mathrm{S18}}}$ & $\mathcal{U}(0, 5)$ & $2.89^{+0.74}_{-0.85}$ \\
$\log{\Delta Q_{\mathrm{S18}}}$ & $\mathcal{U}(-10, 10)$ & $-4.0^{+3.3}_{-4.8}$ \\
$f_{\mathrm{GP,S18}}$ & $\mathcal{U}(0, 1)$ & $0.035^{+0.019}_{-0.032}$ \\
$\sigma_{\mathrm{jit,S18}}$ (ppt) & $\mathcal{LU}(10^{-3}, 10)$ & $0.015^{+0.024}_{-0.014}$ \\
$D_{\mathrm{S18}}$ & $\mathcal{U}(0.5, 1.5)$ & $1.060^{+0.017}_{-0.018}$ \\
\multicolumn{3}{l}{\textbf{TESS Sector 58 Photometry}} \\
$F_{0,\mathrm{S58}}$ & $\mathcal{U}(-0.1, 0.1)$ & $-0.00192^{+0.00031}_{-0.00034}$ \\
$\sigma_{\mathrm{GP,S58}}$ (ppt) & $\mathcal{LU}(0.1, 50)$ & $6.6^{+1.5}_{-2.2}$ \\
$P_{\mathrm{GP,S58}}$ (days) & $\mathcal{U}(4, 20)$ & $5.367^{+0.093}_{-0.089}$ \\
$\log{Q_{0,\mathrm{S58}}}$ & $\mathcal{U}(0, 5)$ & $3.08^{+0.52}_{-0.66}$ \\
$\log{\Delta Q_{\mathrm{S58}}}$ & $\mathcal{U}(-10, 10)$ & $-3.6^{+3.1}_{-5.6}$ \\
$f_{\mathrm{GP,S58}}$ & $\mathcal{U}(0, 1)$ & $0.17^{+0.08}_{-0.11}$ \\
$\sigma_{\mathrm{jit,S58}}$ (ppt) & $\mathcal{LU}(10^{-3}, 10)$ & $0.010^{+0.012}_{-0.009}$ \\
$D_{\mathrm{S58}}$ & $\mathcal{U}(0.5, 1.5)$ & $1.001^{+0.017}_{-0.014}$ \\
\multicolumn{3}{l}{\textbf{TESS Sector 85 Photometry}} \\
$F_{0,\mathrm{S85}}$ & $\mathcal{U}(-0.1, 0.1)$ & $0.00028^{+0.00081}_{-0.00093}$ \\
$\sigma_{\mathrm{GP,S85}}$ (ppt) & $\mathcal{LU}(0.1, 50)$ & $8.2^{+1.2}_{-1.7}$ \\
$P_{\mathrm{GP,S85}}$ (days) & $\mathcal{U}(4, 20)$ & $5.55^{+0.22}_{-0.20}$ \\
$\log{Q_{0,\mathrm{S85}}}$ & $\mathcal{U}(0, 5)$ & $0.099^{+0.064}_{-0.099}$ \\
$\log{\Delta Q_{\mathrm{S85}}}$ & $\mathcal{U}(-10, 10)$ & $6.8^{+3.0}_{-1.4}$ \\
$f_{\mathrm{GP,S85}}$ & $\mathcal{U}(0, 1)$ & $0.73^{+0.27}_{-0.14}$ \\
$\sigma_{\mathrm{jit,S85}}$ (ppt) & $\mathcal{LU}(10^{-3}, 10)$ & $0.008^{+0.011}_{-0.007}$ \\
$D_{\mathrm{S85}}$ & $\mathcal{U}(0.5, 1.5)$ & $1.052^{+0.018}_{-0.020}$ \\
\multicolumn{3}{l}{\textbf{RV Instrumental Parameters}} \\
$\sigma_\mathrm{jit,MX\text{-}Blue\text{-}1}$ (m s$^{-1}$) & $\mathcal{LU}(0.1, 100)$ & $7.6^{+3.0}_{-7.5}$ \\
$\gamma_\mathrm{MX\text{-}Blue\text{-}1}$ (m s$^{-1}$) & $\mathcal{U}(-100, 100)$ & $12.4^{+3.5}_{-2.3}$ \\
$\sigma_\mathrm{jit,MX\text{-}Blue\text{-}2}$ (m s$^{-1}$) & $\mathcal{LU}(0.1, 100)$ & $1.8^{+3.5}_{-1.7}$ \\
$\gamma_\mathrm{MX\text{-}Blue\text{-}2}$ (m s$^{-1}$) & $\mathcal{U}(-100, 100)$ & $35.0^{+3.7}_{-3.2}$ \\
$\sigma_\mathrm{jit,MX\text{-}Red\text{-}1}$ (m s$^{-1}$) & $\mathcal{LU}(0.1, 100)$ & $1.9^{+2.9}_{-1.8}$ \\
$\gamma_\mathrm{MX\text{-}Red\text{-}1}$ (m s$^{-1}$) & $\mathcal{U}(-100, 100)$ & $10.0^{+3.4}_{-2.4}$ \\
$\sigma_\mathrm{jit,MX\text{-}Red\text{-}2}$ (m s$^{-1}$) & $\mathcal{LU}(0.1, 100)$ & $1.7^{+2.6}_{-1.6}$ \\
$\gamma_\mathrm{MX\text{-}Red\text{-}2}$ (m s$^{-1}$) & $\mathcal{U}(-100, 100)$ & $15.7^{+3.8}_{-3.3}$ \\
$\sigma_\mathrm{jit,NEID\text{-}3}$ (m s$^{-1}$) & $\mathcal{LU}(0.1, 100)$ & $11^{+10}_{-11}$ \\
$\gamma_\mathrm{NEID\text{-}3}$ (m s$^{-1}$) & $\mathcal{U}(-12000, -11900)$ & $-11954.1^{+1.5}_{-1.1}$ \\
\multicolumn{3}{l}{\textbf{RV Activity GP}} \\
$\sigma_{\mathrm{RV,GP}}$ (m s$^{-1}$) & $\mathcal{U}(0.05, 50)$ & $30.7^{+5.8}_{-6.7}$ \\
$\log{Q_{0,\mathrm{RV}}}$ & $\mathcal{U}(0, 20)$ & $2.0^{+1.2}_{-1.4}$ \\
$\log{\Delta Q_{\mathrm{RV}}}$ & $\mathcal{U}(-10, 20)$ & $-2.9^{+5.0}_{-3.8}$ \\
$f_{\mathrm{GP,RV}}$ & $\mathcal{U}(0, 1)$ & $0.36^{+0.17}_{-0.30}$%
\enddata
\tablecomments{$\mathcal{U}(a,b)$, $\mathcal{LU}(a,b)$ and $\mathcal{N}(\mu,\sigma)$ denote uniform, log-uniform and Gaussian priors. The posterior column reports the median and 68\% highest-density intervals for each parameter. All orbital elements are osculating elements at the beginning of the integration, taken to be 0.1 days before the first observed transit of TOI-3686\,c, $T_\mathrm{ref} = 2458792.557987$.
}
\tablenotetext{a}{We sampled eccentricity in terms of $\ecosw, \esinw$, which imposes an implicit linear prior on $e$ \citep{Ford2006} due to the Jacobian of the coordinate transformation. We include a factor of $\ln e$ in the posterior evaluation to achieve a uniform prior on $e$.}
\tablenotetext{b}{Limb-darkening coefficients were parameterized in terms of ($q_1, q_2$) as defined by \citet{limbdarkening_Kipping2013}.}
\end{deluxetable*}

\section{Dynamics \& Formation of TOI-3686 \label{sec:dynamics}}

\begin{figure*}
    \includegraphics[width=\linewidth]{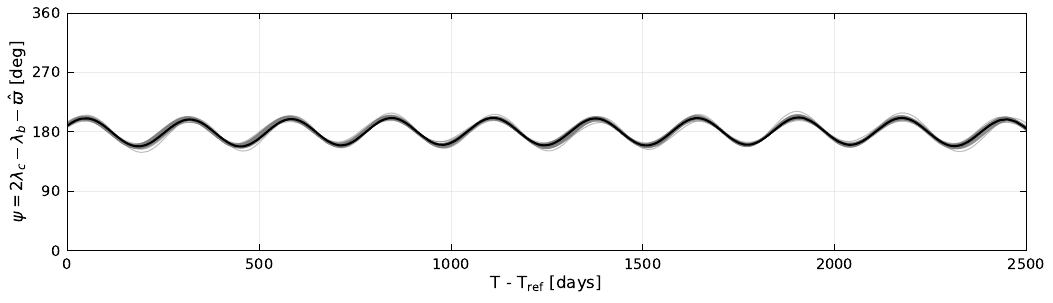}
    \includegraphics[width=0.49\linewidth]{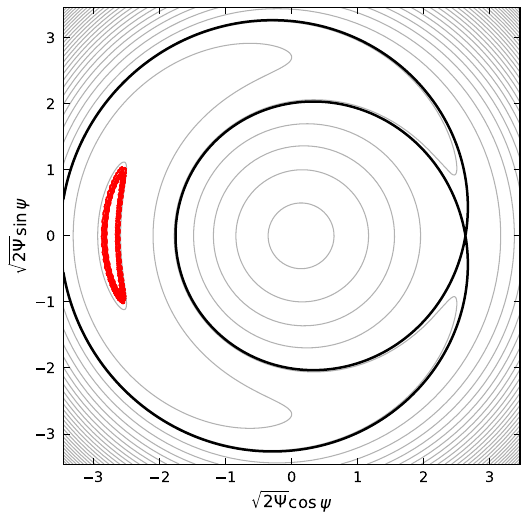}
    \includegraphics[width=0.49\linewidth]{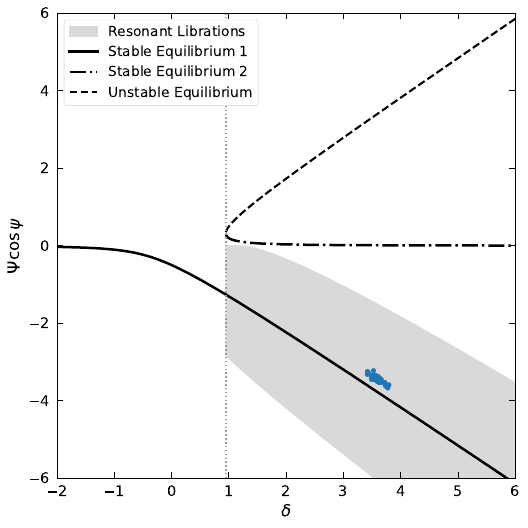}
\caption{\textbf{Top:} Time-evolution of the mixed resonant angle, showing libration around 180$^\circ$. The black line shows the MAP solution and thin gray lines show 50 random posterior draws.
\textbf{Bottom left:} Phase portrait showing the system's resonant trajectory (red) in the canonical coordinates $\sqrt{2\Psi}\sin{\psi}$, $\sqrt{2\Psi}\cos{\psi}$ from the 1-dimensional resonance model of \citet{Nesvorny2016}. Gray lines are level curves of the Hamiltonian, showing two stable equilibria, while the solid black line marks the resonant separatrix.
\textbf{Bottom right:} Fixed points of the same model as a function of proximity to resonance, $\delta$. Resonant trajectories exist only for $\delta > 0.945$ (vertical dotted line), with librations occurring within the gray shaded region. Blue points represent the plotted quantities measured from $N$-body integrations, using 50 random initial conditions drawn from the posterior.
Tidal damping of the inner planet's eccentricity would cause the system to evolve along the solid line, eventually taking the system out of resonance.
\label{fig:resonance}
}
\end{figure*}

We investigate the resonant dynamics of the two planets in the TOI-3686 system.
The top panel of Figure\,\ref{fig:resonance} shows the time evolution of the mixed resonant angle, $\psi = 2\lambda_c - \lambda_b-\hat{\varpi}_{bc}$. Here, $\lambda_b$ and $\lambda_c$ are the planets' mean longitudes, while $\hat{\varpi}_{bc} = \mathrm{Arg}(f e_b e^{i\varpi{_b}} + g e_c e^{i\varpi{_c}})$ is the generalized longitude of pericenter \citep[e.g.,][]{Hadden2019}, and $f, g$ are functions of Laplace coefficients \citep[e.g., Table 3 in][]{Lithwick2012b}.
We find that $\psi$ librates around 180$^\circ$, as expected for a system in an MMR, with an amplitude $A_\psi \sim 20^\circ$ and a period $P_\psi \sim 260$~days; it is the libration of this angle that drives the primary TTV signal for systems in resonance.

We transformed the orbital elements to the 1-dimensional resonance model derived by \citet{Nesvorny2016}, with the following Hamiltonian:
\begin{equation}
    H = -(\Psi-\delta)^2 - \sqrt{2\Psi}\cos{\psi}.
\end{equation}
Here, $\Psi$ is the action conjugate to the mixed resonant angle $\psi$ and is a function of the planets' masses, eccentricities, and semimajor axes.
The parameter $\delta$ denotes the system's proximity to resonance and differs from the distance from period commensurability $\Delta$ (Equation\,\ref{eq:Delta}), as $\delta$ depends on the planets' period ratios, masses, and eccentricities. 
Resonance requires $\delta \gtrsim 0.945$, as no separatrix exists for lower values of $\delta$.
For the TOI-3686 system, we find $\delta = 3.64$, indicating that the system is deep in resonance (lower panels of Figure \ref{fig:resonance}).
Capture into MMR is a natural outcome of migration within a protoplanetary disk \citep{Nelson2002,Lee2002,Kley2004}, thus one plausible formation history of the TOI-3686 planets is that they arrived at their present locations after migrating through the disk.

This resonant configuration differs from other known hot Jupiter systems, even those with nearby transiting companions (Figure \ref{fig:family_portrait}).
While the TOI-1130 \citep{TOI1130_Huang2020b,TOI1130_Korth2023,TOI1130_Borsato2024} and TOI-1408 \citep{TOI1408_Korth2024} systems contain hot Jupiters and inner companions with near-integer period ratios and large TTVs, both have $\delta < 0.945$ and are formally outside the resonant regime.
Instead, this configuration calls to mind the TOI-216 system \citep{TOI216_Dawson2019,TOI216_Dawson2021,TOI216_Nesvorny2022}, which contains an inner Neptune-mass super-puff ($P_b = 17.2$~days) and an outer warm Jupiter ($P_c = 34.5$~days) that are also in a resonant state ($\delta = 2.1$).
Interestingly, while being at longer orbital periods, the TOI-216 planets are similar to TOI-3686 in mass ($M_b = 18\,\Mearth$, $M_c = 0.53\,\Mjup$; \citealt{McKee2023}) and eccentricity ($e_b = 0.16$, $e_c \sim 0$). 

Capture into MMR requires that the two planets undergo convergent migration, such that their period ratio approaches integer commensurability from above \citep{Lee2002,Batygin2015}.
However, the outer planet TOI-3686\,c is likely massive enough to have opened a gap in the disk leading to slower type II migration, so the two planets may have undergone divergent migration, depending on disk properties \citep{Kanagawa2020}.
In the case of TOI-216, \citet{TOI216_Nesvorny2022} argued that the $\Porb = 17$~day inner Neptune was trapped in the transition region between the disk's inner magnetorotational instability (MRI)-active zone and the outer dead zone, allowing the outer planet's migration to catch up and capture the planet pair into resonance.
Given the smaller orbital separation of TOI-3686\,b ($a/\Rstar = 7.2\pm0.1$), its inward migration may instead have stalled at the inner disk edge. For a young, solar-mass star, the magnetospheric truncation radius is expected to be $\sim 7\,\Rstar$ \citep{Bouvier2007}, precisely where we observe TOI-3686\,b today.
The planets in the TOI-3686 system are thus consistent with having formed further out in the disk and migrated inward, with trapping at the disk edge enabling resonant capture and setting the innermost orbital period at a location similar to planets in other exoplanetary systems \citep[e.g.,][]{Lee2017,Batygin2023}.

Given its short orbital period, planet b's moderate eccentricity ($e_b = 0.16$) is surprising as tides raised on the planet by its host star should act to circularize its orbit.
Assuming the constant phase lag model of equilibrium tides, its eccentricity should be damped on a timescale of \citep[e.g.,][]{Jackson2008}:
\begin{equation}
\tau_{e} = \frac{4}{63}\frac{Q^\prime_b}{n_b}\frac{M_b}{M_\star}\left(\frac{a_b}{R_b}\right)^5 \approx 0.6\,\left(\frac{Q^\prime_b}{10^5}\right)\,\mathrm{Gyr}.
\label{eq:tau_e}
\end{equation}
This eccentricity damping should lead to resonant repulsion, with the planet's semimajor axis shrinking at the rate of $\dot{a}/a = -2e^2/\tau_e$ \citep{Papaloizou2010,Lithwick2012a,Batygin2013}.
We ran a long-term forward integration of the system using the \texttt{rebound} code \citep{rebound_Rein2012}, incorporating a dissipative eccentricity damping force through \texttt{reboundx} \citep{reboundx_Tamayo2020}.
These integrations showed that within a time of $0.3\times\tau_e$, the orbit of TOI-3686\,b would have decayed sufficiently that the system is no longer in the resonant regime, with $\delta < 0.945$.

The resonance could be preserved if the hot Jupiter's orbit is shrinking more quickly than that of the inner planet, counteracting the resonant repulsion.
Because the host star is presently spinning more slowly than the orbital period of TOI-3686\,c, tides raised on the host star would shrink the planet's orbit \citep{Rasio1996a}, on a timescale of
\begin{equation}
\tau_{a,\mathrm{tides}} = \frac{2}{9}\,\frac{Q^\prime_\star}{n}\,\frac{M_\star}{M_p}\left(\frac{a}{R_\star}\right)^5.
\end{equation}
Estimates for the stellar reduced tidal quality factor $Q^\prime_\star$ vary by several orders of magnitude \citep[e.g.,][]{Meibom2005}, but even using the relatively high dissipation rate of $Q^\prime_\star = 2\times10^5$ measured from the orbital decay of WASP-12b \citep{Yee2020}, we find an orbital decay timescale for TOI-3686\,c of $\tau_{a,c} = 37$~Gyr, too slow to counterbalance the shrinking of planet b's orbit due to resonant repulsion. In fact, stellar tides alone would also cause the inner planet's orbit to decay faster than the outer planet, with $\tau_{a,b}/\tau_{a,c} = 0.85$.
Thus, tidally-driven orbital decay cannot preserve the system's resonant state over timescales much longer than $\tau_e$.

A more plausible explanation is that the current dynamical state of TOI-3686 is only a temporary phase of the system's evolution, and it will eventually relax toward more circular, nonresonant orbits.
As Equation\,\ref{eq:tau_e} indicates, a planetary reduced tidal quality factor of $Q^\prime_b \sim 10^5$ implies an eccentricity damping timescale of 0.6~Gyr, comparable to the system's gyrochronological age (\S\ref{ssec:youth}).
This value is somewhat less dissipative than the $10^2$--$10^4$ typically assumed for sub-Neptunes, but is consistent with recent estimates of this quantity---for example, \citet{Wang2026b} recently suggested $Q^\prime>10^5$ to explain the remnant eccentricities in the TOI-4495 system, which contains two near-resonant sub-Neptunes.

If TOI-3686 is indeed young and still undergoing tidal circularization and resonant repulsion, this could explain why it is the only one of the known hot Jupiter systems with transiting companions to be in a resonant state.
The other such systems all have older ages, and could simply have had more time to evolve out of resonance.
Indeed, \citet{TOI1130_Borsato2024} attributed the location of the TOI-1130 planets just outside the formal resonant domain to past tidal evolution.
TOI-3686 may therefore be an early snapshot of an evolutionary sequence in which hot Jupiters and their inner companions are captured into resonance during disk migration and then gradually pushed out of it by tides, leaving behind the near-resonant or non-resonant pairs seen around older stars.
A similar trend has been observed for compact multi-planet systems containing smaller planets, which appear to be more commonly found to be in or close to MMR around young stars, with their frequency declining over 100~Myr to Gyr timescales \citep{Hamer2024,Dai2024}.

\section{A Size Dichotomy Between Hot Jupiters and their Close Friends \label{sec:context}}

\begin{figure}
\centering
\includegraphics[width=\linewidth]{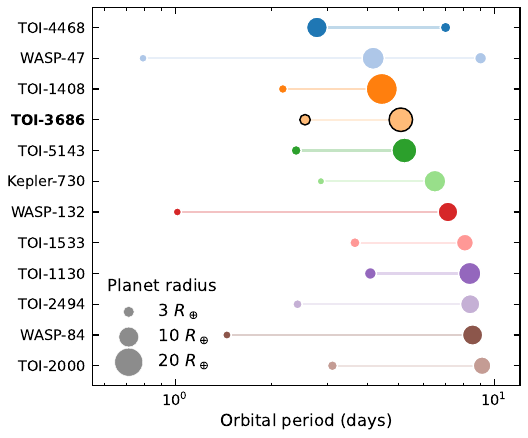}
\includegraphics[width=\linewidth]{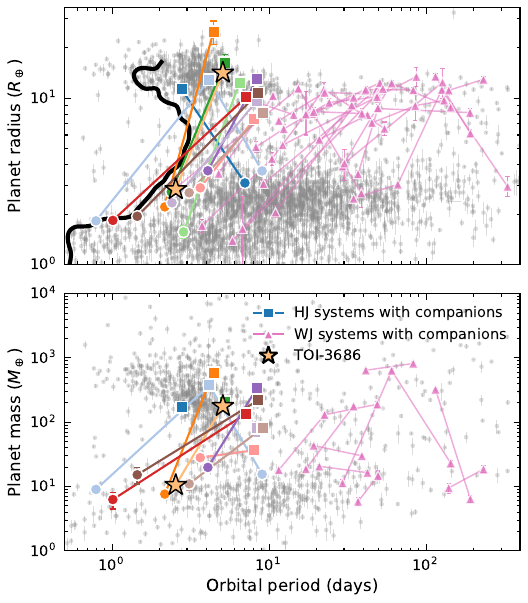}
\caption{\textbf{Top:} All hot Jupiters ($\Porb < 10$~days, $\Rp > 7\,\Rearth$) with known transiting companions. Circles are proportional to the planet radii, with values from the NASA Exoplanet Archive \citep[NEA,][]{NEA_ExoFoP_Christiansen2025}.
\textbf{Middle and bottom:} The same systems in period-radius and period-mass space, with hot Jupiters as squares, and their companions as circles.
For comparison, pink triangles show warm Jupiters ($10 < \Porb < 100$~days) with companions, while gray points show other planets in the NEA with radii or masses measured to better than 3$\sigma$ in their respective panels. 
In the middle panel, the black line shows the ``Neptunian desert'' boundary derived by \citet{Castro-Gonzalez2026} using a Gaussian kernel density estimation model of the known exoplanet population.
\label{fig:family_portrait}}
\end{figure}

Figure \ref{fig:family_portrait} shows TOI-3686 alongside the other known systems containing a hot Jupiter ($\Rp > 7\,\Rearth$, $\Porb < 10$~days) and a transiting companion.\footnote{We have excluded the HIP 67522 system, as it is only 17~Myr old and its planets are probably inflated sub-Neptunes, rather than true hot Jupiters \citep{HIP67522c_Barber2024}.}
An interesting pattern that can be observed from the growing number of such systems is that they exhibit a strong size dichotomy, with the companions to the hot Jupiter being significantly smaller in size.
This disparity stands in contrast to the ``peas-in-a-pod'' pattern observed for compact multi-planet systems of super-Earths and sub-Neptunes \citep{Millholland2017a,Weiss2018}, where planets within the same system tend to have similar sizes.
We performed a Kolmogorov-Smirnov test comparing the distribution of radius ratios between hot Jupiters and their companions with the same distribution for the compact multi-planet systems in \citet{Weiss2018}, finding that they are clearly drawn from different distributions ($p = 8\times10^{-8}$).

The radius contrast for hot Jupiters with companions also stands apart from systems containing a longer-period warm Jupiter ($10\,\mathrm{days}<\Porb < 100\,\mathrm{days}$) and another transiting planet, as seen in Figure \ref{fig:family_portrait}. While there are systems containing one warm Jupiter and a much smaller planet, there are also many examples of systems containing warm Jupiters with additional transiting giant planets.
In contrast, there are no known hot Jupiter systems containing multiple transiting giant planets.\footnote{One near-exception is the WASP-148 system, which contains a hot Jupiter with a non-transiting outer giant planet \citep{WASP148_Hebrard2020}.}
A Kolmogorov-Smirnov test comparing the radius ratios between hot and warm Jupiters with companions is marginally significant, with $p = 0.061$.

A connected observation is that companions to hot Jupiters are typically among the largest and most massive of the super-Earth/sub-Neptune population for their orbital period, with a median companion mass of $11\,\Mearth$, close to the $\sim 10\,\Mearth$ threshold thought to be necessary for initiating runaway gas accretion \citep{Pollack1996}.
One possible interpretation then is that these systems may have initially formed as pairs of planets with similar masses, but only one subsequently accreted a substantial gas envelope, leading to a stark size dichotomy.
Further enhancement of the size and mass contrast for systems containing the shortest-period planets may then subsequently occur due to photoevaporation \citep{Owen2013,Lopez2013}, as some of these lie on the lower edge of the ``Neptunian desert'' \citep{Szabo2011,Beauge2013,Mazeh2016,Castro-Gonzalez2024}.
The difference between the hot and warm Jupiters with companions may be a consequence of the fact that the warm Jupiter systems are less tightly-packed, potentially allowing multiple cores to grow into giant planets.

One caveat to this interpretation is that only eight of the twelve known transiting companions to hot Jupiters have measured masses, and both the RV and TTV techniques are more sensitive to more massive planets.
Detection with the transit method is also biased toward larger radii, and most of the known companions, including TOI-3686\,b, lie close to TESS' detection limits.
Mass measurements for the remaining systems, as well as more sensitive transit surveys like PLATO \citep{PLATO_Rauer2025} that could detect smaller planets, would test whether companions to hot Jupiters are systematically larger and more massive than typical sub-Neptunes.
If so, these systems may be the high-mass extremes of the processes that form other compact multi-planet systems, albeit where one planet crossed the threshold for undergoing runaway gas accretion.

\section{Conclusions \label{sec:conclusions}}

We have presented the discovery and characterization of the TOI-3686 system.
The 40-minute TTVs of TOI-3686\,c are amongst the largest ever observed for a hot Jupiter, while the 10~hour TTVs of the inner sub-Neptune, have the third largest TTV amplitude of any planet measured relative to its orbital period, after TOI-1408\,c \citep{TOI1408_Korth2024} and TOI-4504\,d \citep{TOI4504_Vitkova2024,TOI4504_Almenara2026}.

Dynamical modeling of TESS and extensive ground-based data reveal that the two planets are deep in a 2:1 MMR.
TOI-3686 joins the handful of hot Jupiters known to have nearby transiting companions, and is the only such system known to be in a truly resonant state.
Because tidal dissipation is expected to eventually drive the system out of resonance, that we are observing the system in this resonant configuration implies a reduced tidal quality factor for the inner sub-Neptune of $Q^\prime_p \gtrsim 10^5$ for its eccentricity to have been maintained over the $\lesssim1$~Gyr system age.

TESS' all-sky transit survey and growing observation baseline is enabling it to discover new, rare examples of non-isolated hot Jupiters, by detecting the transits of or TTVs induced by the companions to hot Jupiters.
Such systems trace a dynamically quiescent pathway for producing close-in giant planets, likely involving early migration within the disk, and may represent the high mass end of planet formation processes that produce compact, multi-planet systems of smaller planets.
More examples of hot Jupiters with significant TTVs continue to be discovered (e.g., \citealp{TOI1408_Korth2024,TOI1130_Borsato2024}; S.~W.~Yee et al.~accepted; S.~Stefanov et al.~submitted), adding to the benchmark sample of hot Jupiters that are in or close to MMRs; these will be valuable systems that may help shed light on how giant planets form and interact with their protoplanetary disks.

\begin{acknowledgments}
S.W.Y. gratefully acknowledges support from the Heising-Simons Foundation. T.I.D. acknowledges support from the Harvard Origins of Life initiative. 
K.A.C. acknowledges support from the TESS mission via subaward s3449 from MIT and NASA grants 80NSSC24K1889 and 80NSSC26K0081. E.N. acknowledges support from the TESS mission via NASA grants 80NSSC24K1889 and 80NSSC26K0081.
Funding for K.B. was provided by the European Union (ERC AdG SUBSTELLAR, GA 101054354).

The authors thank the observers at the WIYN 3.5m and Gemini-North telescopes for their skillful execution of the observations, and the NEID and MAROON-X instrument teams for providing the reduced data products.

AI tools (ChatGPT 5.6 Sol, Claude Sonnet 5.0, Claude Opus 5.5) were used to improve the visualizations in this manuscript. No text in this manuscript was generated by LLMs. Responsibility for the final manuscript lies entirely with the authors.

This paper includes data collected by the TESS mission, which are publicly available from the Mikulski Archive for Space Telescopes (MAST): \dataset[doi: 10.17909/8ps9-1y38]{\doi{10.17909/8ps9-1y38}}. Funding for the TESS mission is provided by the NASA's Science Mission Directorate. We acknowledge the use of public TOI Release data from pipelines at the TESS Science Office and at the TESS Science Processing Operations Center. 
This research has made use of the Exoplanet Follow-up Observation Program (ExoFOP; DOI: 10.26134/ExoFOP5) website, which is operated by the California Institute of Technology, under contract with the National Aeronautics and Space Administration under the Exoplanet Exploration Program.

Part of the data presented herein were obtained with the WIYN 3.5m from telescope time allocated under the NN-EXPLORE program, with support from the National Aeronautics and Space Administration and the U.S. National Science Foundation. Some of the data were obtained with the NEID spectrograph built by Penn State University and operated at the WIYN Observatory by NSF NOIRLab, under the NN-EXPLORE partnership of the National Aeronautics and Space Administration and the U.S. National Science Foundation. The authors are honored to be permitted to conduct astronomical research on I'oligam Du'ag (Kitt Peak), a mountain with particular significance to the Tohono O'odham.

This work makes use of observations from the LCOGT network. Part of the LCOGT telescope time was granted by NOIRLab through the Mid-Scale Innovations Program (MSIP). MSIP is funded by NSF. 

This article is based on observations made in the 
Two-meter Twin Telescope (TTT)\footnote{\url{http://ttt.iac.es}} sited at the Teide Observatory of the Instituto 
de Astrofísica de Canarias (IAC), that Light Bridges operates in Tenerife, Canary Islands (Spain). The observation 
time rights (DTO) used for this research were consumed in the PEI "SBSTLLAR25". This research used storage and 
computing capacity in ASTRO POC's EDGE computing center at Tenerife under the form of Indefeasible Computer Rights 
(ICR). The ICR were consumed in the PEI “SBSTLLAR25" with the collaboration of Betchle AG.

This work is based on observations obtained at the international Gemini Observatory, a program of NSF NOIRLab, which is managed by the Association of Universities for Research in Astronomy (AURA) under a cooperative agreement with the U.S. National Science Foundation on behalf of the Gemini Observatory partnership: the U.S. National Science Foundation (United States), National Research Council (Canada), Agencia Nacional de Investigaci\'{o}n y Desarrollo (Chile), Ministerio de Ciencia, Tecnolog\'{i}a e Innovaci\'{o}n (Argentina), Minist\'{e}rio da Ci\^{e}ncia, Tecnologia, Inova\c{c}\~{o}es e Comunica\c{c}\~{o}es (Brazil), and Korea Astronomy and Space Science Institute (Republic of Korea). This work was enabled by observations made from the Gemini North telescope, located within the Maunakea Science Reserve and adjacent to the summit of Maunakea. We are grateful for the privilege of observing the Universe from a place that is unique in both its astronomical quality and its cultural significance.

We acknowledge financial support from the Agencia Estatal de Investigaci\'on of the Ministerio de Ciencia e Innovaci\'on MCIN/AEI/10.13039/501100011033 and the ERDF “A way of making Europe” through projects PID2021-125627OB-C32 and PID2024-158486OB-C32. This work is supported by the ERC Grant (ERC Advanced Grant SPEAR, GA 101200674).  Funded by the European Union. Views and opinions expressed are however those of the author(s) only and do not necessarily reflect those of the European Union or the  European Research Council Executive Agency (ERCEA). Neither the European Union nor the granting authority can be held responsible for them.

We acknowledge the Marvel consortium (Institute of Astronomy, KU Leuven (PI); Stockholm University; University of Vienna; Instituto de Astrofísica de Canarias; Institute of Space Studies of Catalonia (IEEC), Barcelona; School of Mathematical and Physical Sciences, Macquarie University; University of Warwick; Ruprecht Karl University of Heidelberg) for the telescope time granted on the Marvel Telescope Array in La Palma, Spain.

\end{acknowledgments}

\facilities{TESS, Gemini:Gillett (MAROON-X), WIYN (NEID), FLWO:1.5m (TRES), FLWO:1.2m (KeplerCam), LCOGT 1.0m (Sinistro)}

\software{\texttt{AstroImageJ} \citep{AstroImageJ_Collins17},
          \texttt{astropy} \citep{Astropy13,Astropy18,Astropy2022},  
          \texttt{EXOFASTv2} \citep{ExoFAST_Eastman2013,ExoFASTv2_Eastman19},
          \texttt{exoplanet} \citep{exoplanet_ForemanMackey2021,exoplanet_zenodo},
          \texttt{jaxoplanet} \citep{jaxoplanet_Hattori2024},
          \texttt{jnkepler} \citep{jnkepler_Masuda2024,jnkepler_ASCL_Masuda2025},
          \texttt{numpyro} \citep{numpyro_Bingham2019,numpyro_Phan2019},
          \texttt{Tapir} \citep{TAPIR_Jensen2013}}

\bibliography{main,software,instruments,catalogs}{}
\bibliographystyle{aasjournalv7}

\end{document}